\documentclass[journal,onecolumn,a4paper]{IEEEtran}
\usepackage[utf8]{inputenc}
\usepackage{amsfonts}
\usepackage{amsmath}
\usepackage{amsthm}
\usepackage{amscd}
\usepackage{graphicx}
  \graphicspath{{/}{figures/}}
  \DeclareGraphicsExtensions{.pdf,.png}
\usepackage{epstopdf}
\usepackage{todonotes}
\usepackage{comment}
\usepackage{enumitem}
\usepackage{booktabs}

\newcounter{mycomment}

\SetLabelAlign{parleft}{\parbox[t]{\labelwidth}{\raggedleft\textbf{#1}}}
\setlist[description]{align=parleft,labelindent=0.0ex,labelwidth=7ex}

\newtheorem{Definition}{Definition}
\newtheorem{Theorem}{Theorem}

\newcommand*{\ii}{\ensuremath{\mathrm{i}}}

\newcommand*{\ICHAIN}{\ensuremath{(\mathrm{ITD}-\mathrm{IMF})\mathrm{chain}_{1}}}

\begin{document}

\title{Kolmogorov Space in  Time Series Data}

\author{
\IEEEauthorblockN{K. Kanjamapornkul\IEEEauthorrefmark{1},
R. Pinčák\IEEEauthorrefmark{2}\IEEEauthorrefmark{3} \\
\IEEEauthorblockA{\IEEEauthorrefmark{1}  Department of Computer Engineering, Faculty of Engineering, Chulalongkorn University, 254 Phyathai Road, Bangkok, Thailand, Email: kabinsky@hotmail.com}\\
\IEEEauthorblockA{\IEEEauthorrefmark{2} Institute of Experimental Physics, Slovak Academy of Sciences, Watsonova 47, 043 53 Košice, Slovak Republic}\\
\IEEEauthorblockA{\IEEEauthorrefmark{3} Bogoliubov Laboratory of Theoretical Physics, Joint Institute for Nuclear Research, \\141980 Dubna, Moscow region, Russia, Email: pincak@saske.sk}\\
}}

\maketitle

\begin{abstract}
We provide the proof that the space of time series data is a Kolmogorov space with $T_{0}$-separation axiom using the loop space of time series data. In our approach we define a cyclic coordinate of intrinsic time scale of time series data after empirical mode decomposition. A spinor field of time series data comes from the rotation of data around price and time axis by defining a new extradimension to time series data. We show that there exist hidden eight dimensions in Kolmogorov space for time series data. Our concept is realized as the algorithm of empirical mode decomposition and intrinsic time scale decomposition and it is subsequently used for preliminary analysis on the real time series data.
\end{abstract}


\IEEEpeerreviewmaketitle

\section{Introduction}\label{sec:intro}

In the mesoscopic quantum world of financial market and market microstructure \cite{Almog2015}, the stock equilibrium price and various types of order submissions, characterized by the behavior of traders in orderbook, exists in many parallel states simultaneously and an act of measurement of equilibrium price itself forces the price and time ordering to collapse to a definite state, so called entanglement state \cite{tan2015}. Recently, many physicists and engineers tried to understand a dynamics of systems with modeling of stock market using empirical analysis of stock price \cite{Rinn2015}. The financial market can be  realized as a topological space of underlying financial time series data. The observed time series data from financial market can be classified as nonlinear and non-stationary system in which typical econometric tool of linear regression models such as ARIMA, GARCH and state space model cannot visualize all multiple processes of complex system such as the financial market \cite{Hung2015}. On the other side, the scientists borrowed from signal processing the data-mining tool such as neural network combined with wavelet transformation or they supported vector machines with some extra datamining tools to predict financial time series with the overfitting and prior problems \cite{Xi201457}. They believe that they can overcome prediction problem by finding a good risk factors to let a Bayesian system to learn \cite{Ticknor} or by regression of those risk factors, but they do not realize the main problem connected with the defect of algebraic topological construction of the data-mining tools. The main defect of data-mining tools is based on a fitting problem with one parameter of learning from single stochastic process instead of infinite factors in which the influence of infinite stochastic processes governs on future expectation price. In other words one financial time series are composed of infinitely many random variables in which the average of all random variables not always converge to single Kolmogorov space over Euclidean space of time series data. When we add one point of future price and make a fitting curve using the data-mining tool for regression and state space model of Markov switching regime the coefficient of equation which we used for description of the historical data will update and change the historical path, so it leads to a nonrealistic situation.

All those problems of the prior effect and endeffect of time series data have intrinsic behavior of algebraic defect of topological space underyling time series data in separable $T_{0}$-axiom of Kolmogorov space between price and time. The problem of forecasting arose from a defect in algebraic and geometric construction of space of time series and it has a deep relationship to an empirical analysis problem of  nonstationarity of time series data and volatility clustering phenomena in financial time series data so called stylized fact \cite{fact} and separation of hidden Markov transition probability state in quantum entanglement state with Hopf fibration. In nature of macroeconomic time series model, we assume equilibrium properties of dynamic stochastic model over stochastic process of deterministic dynamical system with many assumptions. The precise definition of Kolmogorov topological space underlying the financial time series model in spinor field can hopefully introduce the better understanding of the macroeconomic models. The suitable algebraic reconstruction of space of time series, possible under a Kolmogorov space concept with consistent separation axiom, could help with analysis of the prior effect and endeffect of time series models. There are some indications that such a concept could be realized as a quotient topological space with a few hidden states in extradimensions of loop space of time series data. It is possible to connect hidden eight states in a Kolmogorov space with the empirically observed characteristic 
correlation structure patterns \cite{Stanley}. 

A Kolmogorov space \cite{Karno} is a  topological space fulfilling the $T_{0}$-separation axiom, in other words, it is a topological space \cite{Janich} in which every pair of distinct points is topologically distinguishable. A space of time series data represented by the topological space with fixed point property was first considered by research group from Slovak Academy of Sciences \cite{Pincak21}.
A fixed point space \cite{Fixpoint2} need not be Hausdorff space necessarily, but it has to satisfy weaker $T_{0}$-separation axiom, it means that all fixed point spaces are Kolmogorov ones \cite{Fixpoint}. For that reason one must to verify a $T_{0}$-separation axiom property for a space of time series data, in order to declare it to be Kolmogorov space.

As we know, no precise mathematical definition exists for time series and financial time series. We only know that time series are the observations ordered in a time (or space). For time series data, we can not use a set to define time series data directly. Since the data can have the same value in a set notation, the same value cannot be separated using the $T_{0}$-separation under discrete topology. Therefore, a pointed space of time series data is not a Kolmogorov space, one needs to define  extradimensions \cite{Pincak20} in time series data as loop space \cite{loopspace} of path lifting for the separation of data under $T_{0}$-separation axiom. 

The real application of a Kolmogorov space of time series data is a directional prediction. We investigated a loop space of time series data of entanglement state of mixed direction between future direction and past direction in time series data. These states are suitable to open a short position or to open a long position in Stock Index Futures market. The ultimate goal of time series prediction is directional prediction. The typical output of a directional prediction is the prediction to up or down (or down with no direction change) with respect to the present value. We tested the performance of our mathematical modeling using the forecasting methodology over nonlinear and nonstationary time series data of stock market price. There exist new tools for data analysis of nonlinear and nonstationary time series data so called Hilbert Huang transformation \cite{Huang} and intrinsic time scale decompostion (ITD) \cite{Frei}. These tools can be simple used together with the artificial neural network (ANN) to predict the direction of stock price \cite{stock}. However both methodologies have severe problems on boundary condition of time series data so called endeffect \cite{end1, end2}.

The paper is organized as follows. In Section~\ref{sec:concept} we specify the basic definition of Kolmogorov space and how the concept of algebraic topology \cite{Massey} is related to data in time series. In Section~\ref{sec:loop} we define a loop space in time series data by using of extradimensions of underlying topological space. In Section \ref{sec:proof} we prove that there exists a time series data in spinor field \cite{fibre} with underlying structure of Kolmogorov space in time series data. In Section \ref{sec:concl} we discuss the result of the proof and consider about a time series in terms of quaternionic projective space \cite{Arnold}. We prove that a new space of time series is a Kolmogorov space with eight hidden dimensions with spin invariant property in time series data by which it relates to quantum entaglement qubit state \cite{Rigetti} in time series data. In Appendix~\ref{app:data} we provide detail of empirical data analysis with cyclic coordinate in financial time series data by using empirical mode decomposition and intrinsic time scale decomposition.

\section{Kolmogorov space and concept of algebraic topology and their relation to a time series data}\label{sec:concept}

\begin{Definition}[Kolmogorov space]
	A Kolmogorov space $X$ is a topological space fullfilling the $T_{0}$-separation axiom such that for any two points $x, y\in X$, there exists an open set $U$ such that $x\in U$ and $y\notin U$ or $y\in U$ and $x\notin U$.
\end{Definition} 

\begin{table}[!t]
		\caption{The relations between the main separation conditions with implications in downward direction.  
			For instance, every $T_{3}$ space is also a $T_{2}$ space, and every preregular space is also a symmetric space. The table is borrowed from \cite{book}.}\label{Kolmogorovspace}
	\begin{center}
	\begin{tabular}{ccc}
		\toprule
		\textbf{metrizable}  & \textbf{psudometrizable} & \textbf{name of space (process) with metric} \\ \midrule
		paracompact and $T_{0}$ & paracompact (partition of unity) & compact spac (integrable)  \\ \midrule
		$T_{4}=$ normal and $T_{1}$& normal and symmetric& Urysohn space 
		(Shinkings)\\ \midrule
		$T_{3.5}=$ Tychonov & completely regular & Tychonov space (gauges, uniformities)\\ \midrule
		$T_{3}=$ regular and separated & regular & regular space (closed neighborhood bases; extension by continuity)\\ \midrule
		$T_{2}=$ Hausdorff & preregular & Hausdorff space (limits are unique up to topological distinishability)\\ \midrule
		$T_{1}=$ Fréchet & symmetric & symmetric space (the closures of points form a partition of $X$)  \\ \midrule
		$T_{0}=$ Kolmogorov & points are topological distinguishable & arbitrary topological space\\ \bottomrule	 
	\end{tabular}
	\end{center}
\end{table}

The most recent work in financial data analysis \cite{Bartosz} is based on a definition of space of time series $X=\{x_{t}\in \mathbb{R}, t\in \mathbb{N}\},$ without separation axiom in the contribution. Let $X$ be a set of ordered points of time series values. We can consider a set of time series data as an object in categories of SET with objects sets and morphisms -- the injection functions between sets. A functor is a transformation of object of time series from categories of SET into another categories of TOP and GROUP. An example of a set of time series is
\begin{equation}
X=\big\{x_{1},x_{2},x_{3},x_{4},\cdots x_{n}\big\},\quad x_{i}\in \mathbb{R}.
\end{equation}
which induces a sequence of measured values
\begin{equation}
x_{1}\rightarrow x_{2}\rightarrow x_{3}\rightarrow \cdots \rightarrow x_{n} 
\end{equation}
and a sequence of discrete time intervals between measurements
\begin{equation}
t_{1}\rightarrow t_{2}\rightarrow t_{3}\rightarrow \cdots \rightarrow t_{n-1}.
\end{equation}
If we consider a sequence of data just a set, we can define a discrete topology on a set of data. For a time series data, we can not use a set to define a time series data directly, since the data can have the same value in a set's notation. E.~g., if we have a time series data with following six numbers of sample data
\begin{equation}
A=\big\{1,2,3,3,3,4\big\},
\end{equation}
the same values can not be separated. If we use set theory to induce a pointed set topology, we will fail to define an open set of data because,
\begin{equation}
A=\big\{1,2,3,3,3,4\big\}=\big\{  1,2,3,4\big\}.
\end{equation}

\begin{Definition}
	Let $A\neq \phi$ be a set. Let $\tau=P(A)$ be the power set of $A$. Then $\tau$ is called the discrete topology on $A$ and $(A,\tau)=(A,P(A))$ the discrete space on $A$, or just a discrete space.
\end{Definition}
\begin{Definition}[Finite discrete topology]
	If $A$ is finite, $\tau=P(A)$ is a finite discrete topology, and $(A,\tau)=(A,P(A))$ is a finite discrete space.
\end{Definition}

\begin{figure}[!t]
	\centering
	\includegraphics[width=0.45\textwidth]{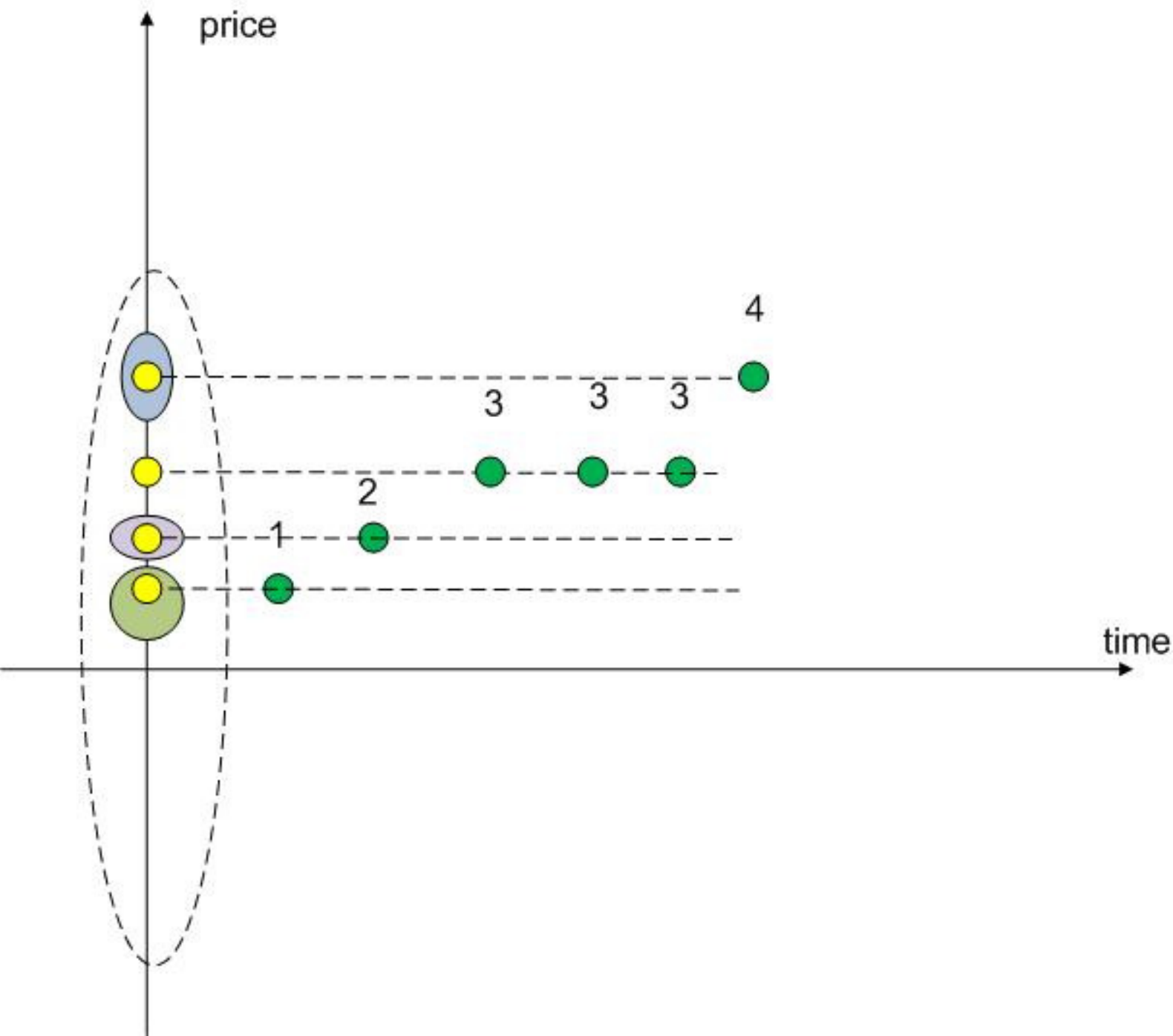}
	\includegraphics[width=0.45\textwidth]{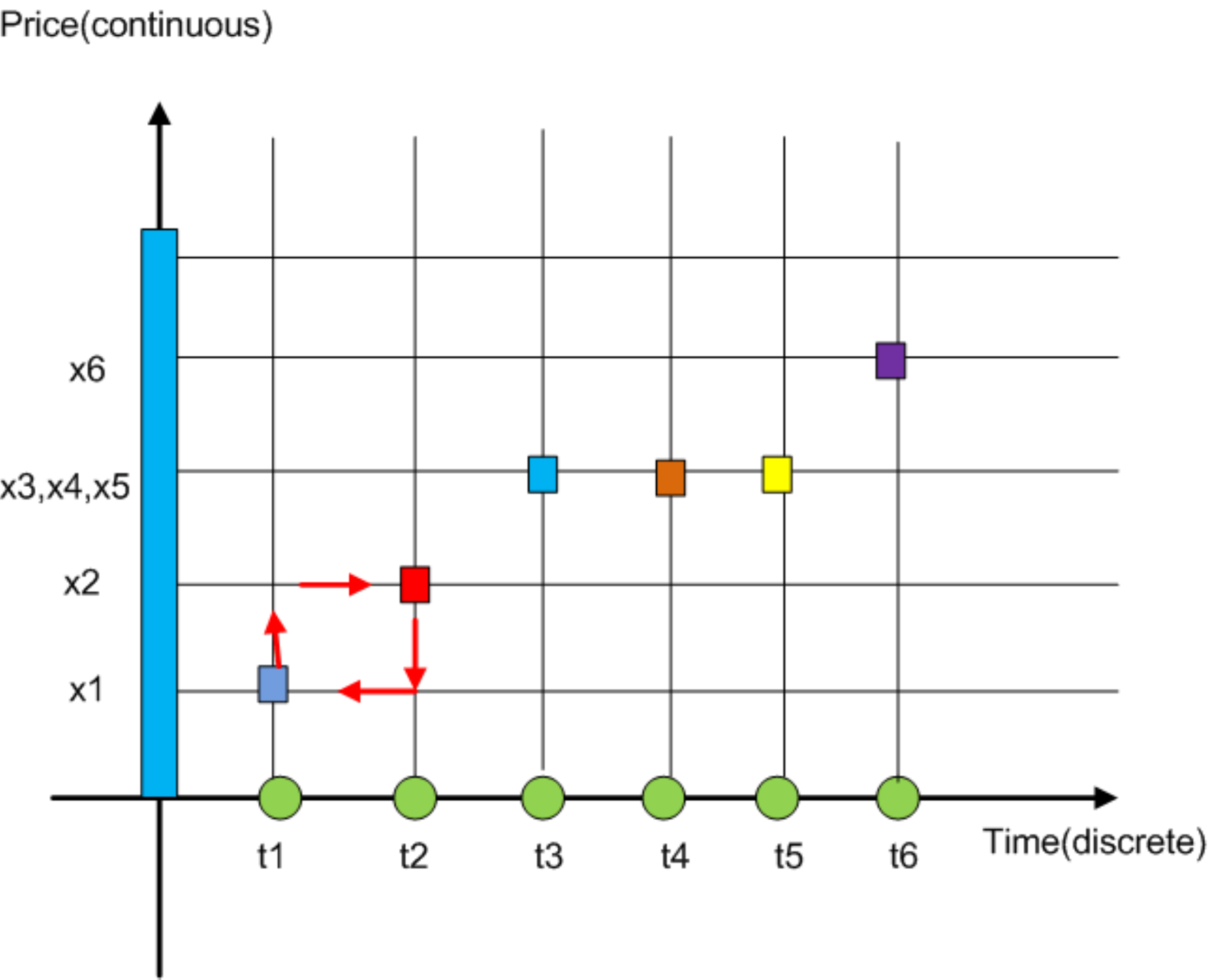}
	\caption{On the left the discrete topology of a time series data with the open set visualized in the projective plane at $y$-axis. We can see that the projection of all three points with the value $3$ is just one point at the same position, we cannot use an open set to separate these three points. The plot on the right shows that the point of time series data need to be embeded in the intersection point of perpendicular line.}	\label{separate}
\end{figure}

\begin{figure}[!t]
	\centering
	\includegraphics[width=0.45\textwidth]{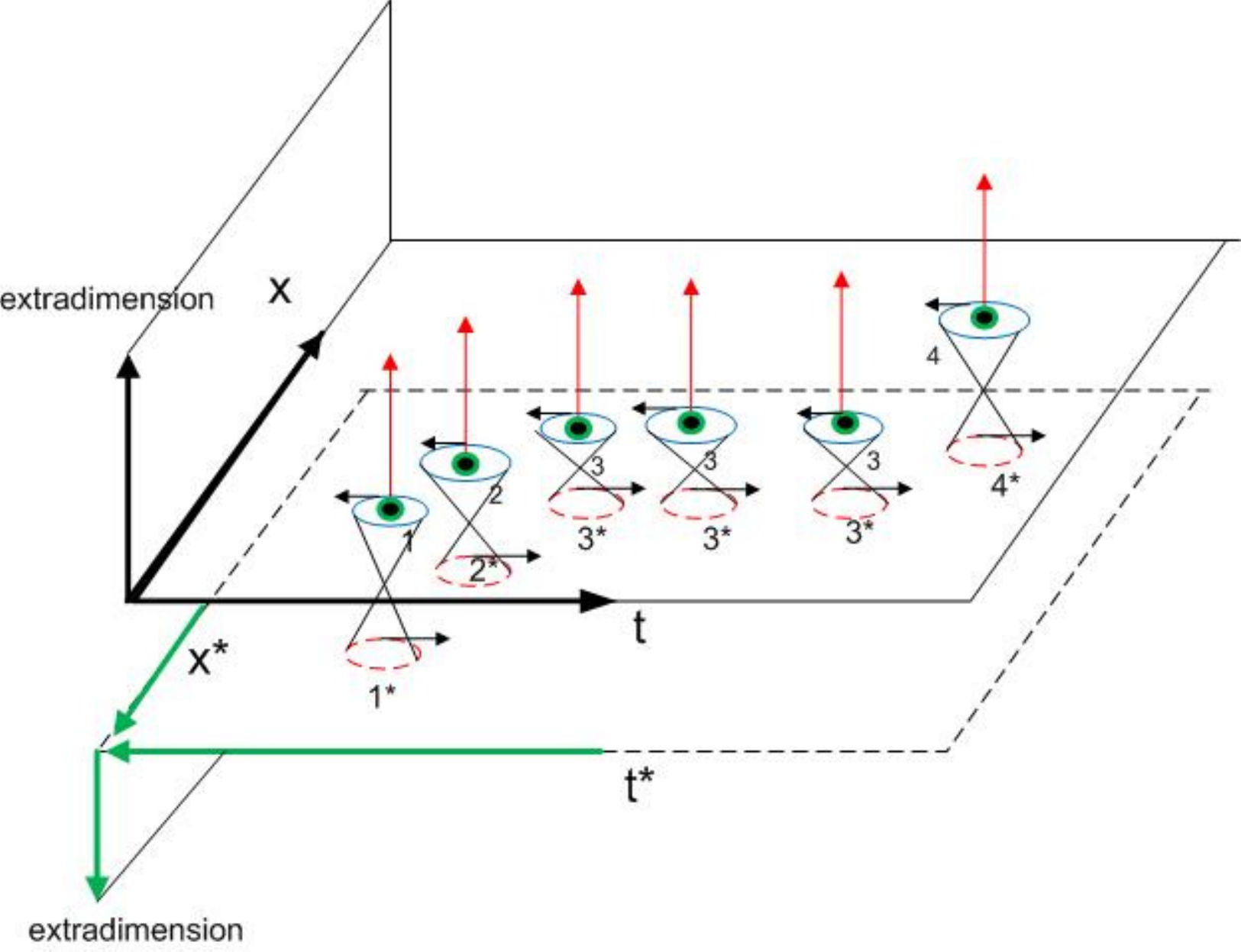}
	\includegraphics[width=0.45\textwidth]{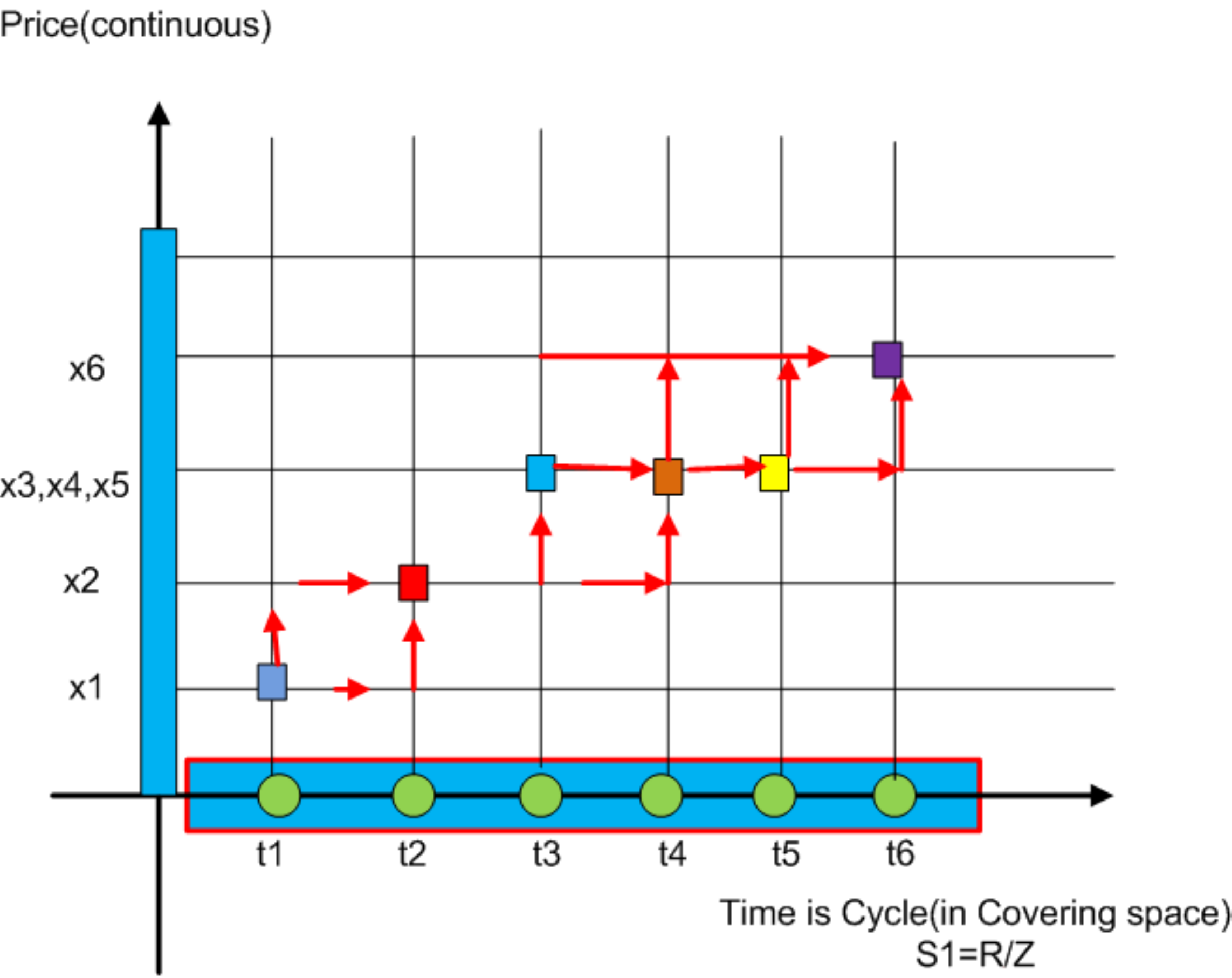}
	\caption{On the left the visualisation of the extradimension for time series. There exists an induced field of cross product between price and time in covariant and contraviant tensor field in hidden plane in extradimension of Euclidean plane of time series data. The right figure shows CW complex of torus. The cell complex is in orientation state, if it is in nonorientation state the glue process will produce Möbius strip of time series data.\label{extradimension}}
\end{figure}

Let a sequence of data be
\begin{equation}
x_{1}\rightarrow x_{2}  \rightarrow x_{3}  \rightarrow x_{4}  \rightarrow x_{5}  \rightarrow x_{6}  
\end{equation}
with the values
\begin{equation}\label{eq:x}
x_{1}=1,\; x_{2}=2,\; x_{3}=3,\; x_{4}=3,\; x_{5}=3,\; x_{6}=4.
\end{equation}
When we use discrete  topology, we will get an open set with $2^{n(A)}=2^{4}=16$ open subsets. A sequence of similar values for $x_{3}=3$, $x_{4}=3$, $x_{5}=3$ will not be separated (Fig.~\ref{separate}),
an open set can not be used to separate all time series data. Therefore with discrete topology a space $(A,\tau)=(A,P(A))$ is a finite discrete space, it is not a Kolmogorov space with $T_{0}$-separation axiom. Let us assume that a time series data is embedded in non-Euclidean plane with extradimension (Fig.~\ref{extradimension}). It allows us to embed a loop structure between all data of time series connected to each other as path components. By this redefinition of a financial time series data, one is allowed to use an equivalent class of loop space (fundamental group of time series) to separate the sequence of similar values for $x_{3}$, $x_{4}$, $x_{5}$ in Eq.~(\ref{eq:x}) by open set in path component.
\begin{figure}[!t]
	\centering
	\includegraphics[width=0.65\textwidth]{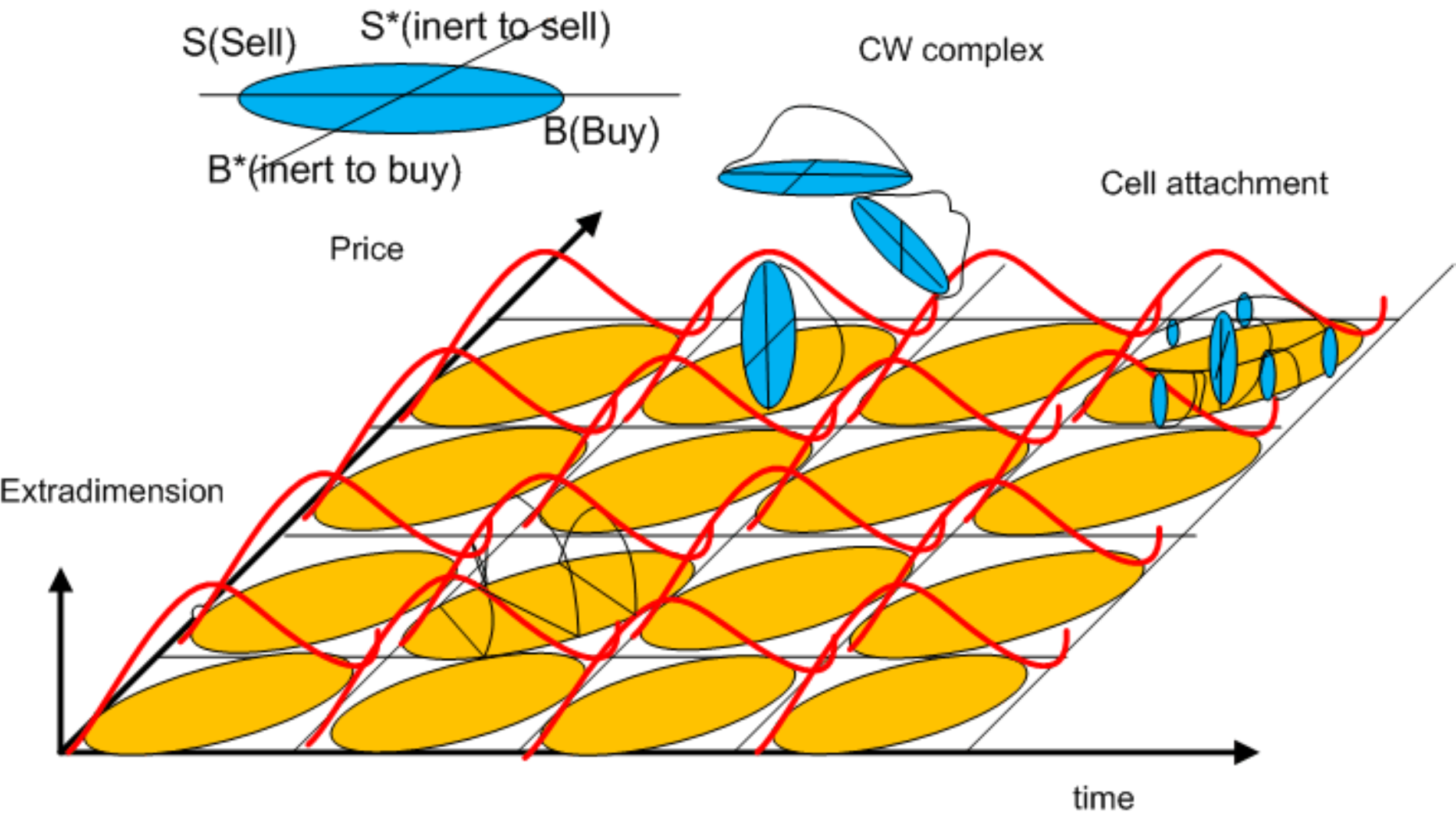}
	\caption{Non-euclidean plane of time series. The demonstration of a mirror symmetry of dual price and dual time scale axis between price and time for covariant and contraviant tensor field in hidden plane in extradimensions of Euclidean plane of time series data.\label{loop1}}
\end{figure}

For time series model a time is discrete. We use algebraic topology tool to change topology of discrete space $\mathbb{N}\subset \mathbb{Z}$ to $\mathbb{R}$ by using quotient topology of covering space $\mathbb{R}/\mathbb{Z}$ with some fibre space. We glue a pointed space of each fibre to form a pointed space of time series data by using coproduct in topology. If we treat time as point in real axis, we will induce an open set. For every  $i=1,\cdots n, t_{i}\in \mathbb{R}$ we will have an open set 
\begin{equation}
(-\infty,t_{1}),(t_{1},t_{2}),(t_{2},t_{3}),\cdots,(t_{n-1},t_{n}),(t_{n},\infty)
\end{equation}
We call this open set a topology for time series data of time path. Let  $t\in A\subset \mathbb{R}$ be an index set of time series data. Let a pointed space of time series data be $X_{t}=\{x_{t}\}$. We define a family of pointed space of time series data by  $\{ X_{t}|t\in A \}$. The family of space underlying measurement data will induce a topological sum of space of time series data $\coprod_{t\in A}  X_{t}$ where
$\cup \{ X_{t}\times \{ t \}|t\in A\}$. We let a closed embedded map
\begin{equation}
i_{\beta}:X_{\beta} \rightarrow \coprod_{t\in A} X_{t},x\mapsto (x,\beta)
\end{equation}
for every $\beta \in A$ we have $i_{\beta}(X_{\beta})\cap i_{\alpha}(X_{\alpha})=\phi$,
if  $\alpha\neq \beta$. 
One of a major problem in financial time series of stock price is how can we incorporate a behavior of trader into financial time series data directly. We solve this problem by using CW decomposition \cite{Janich} of financial time series. We can attach buying and selling operation using a coproduct of topology for each cell decomposition of Euclidean space of time series data (see Fig.~\ref{loop1}). This process induces a hidden dimension in Kolmogorov space of time series data.

\begin{figure}[!t]
	\centering
	\includegraphics[width=0.45\textwidth,angle=0]{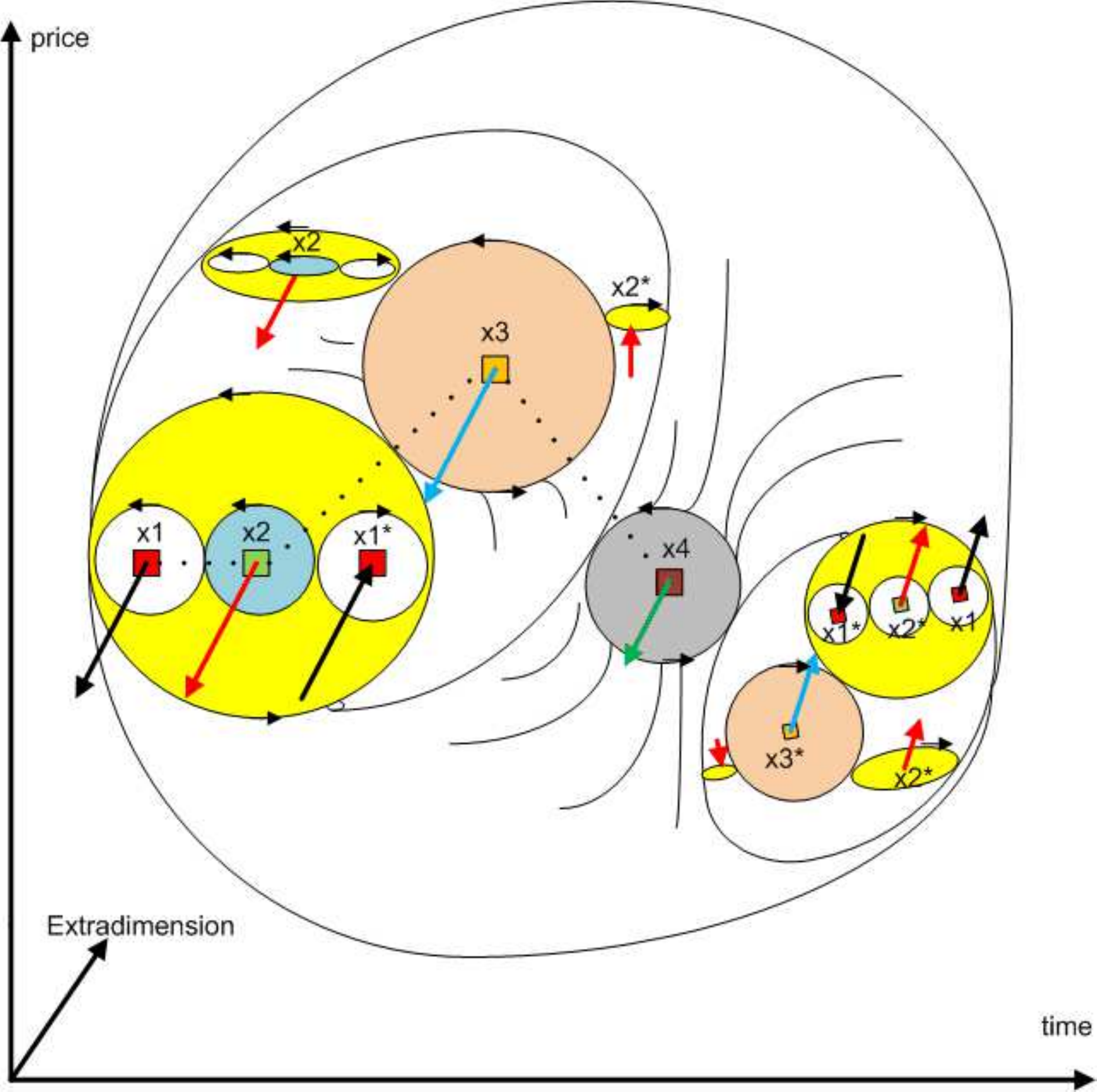}
	\caption{The loop space of time series in which one can separate three same values ``$3$'' with a loop. One has x1$\,=1$, x2$\,=2$, x3$\,=3$, x4$\,=3$, x5$\,=3$, x6$\,=4$ in 3-dimensional blend complex surface model of spinor field of time series data in Kolomogorov space. The thick dark line connecting $x3$ to $x4$ to $x5$ is in a perspective view of 3-dimensional with extradimension add to Euclidean plane by induced equipotential spinor fields of times data of $3, 3, 3$. This fields are propulsion to each other in model of loop space modelling and induce a straight line with equal slope blend to the direction of extradimension (projective of this line is still straight line in Euclidean plane). Therefore values $3, 3, 3$ are completely separated by using $T_{0}$-separation axiom of Kolmogorov space of time series data. \label{separate2}}
\end{figure}

All data which cannot be separated in Euclidean plane (Fig.~\ref{separate2}) now can be separated by using path lifting to open set in covering space of physiology of time series. A ground space of time series data $A$ can be realized as topological space by using disjoint union from
\begin{equation}
A=\big\{1 \big\}\cup   \big\{2 \big\}  \cup \big\{3 \big\} \cup \big\{3 \big\} \cup\big\{3 \big\} \cup \big\{4 \big\}
\end{equation}
to a space of time series $X$,
\begin{equation}
X=\big\{1 \big\}\coprod   \big\{2 \big\}  \coprod  \big\{3 \big\} \coprod  \big\{3 \big\} \coprod \big\{3 \big\} \coprod  \big\{4 \big\}.
\end{equation}

Let $t\in I=[0,1]$ be a time interval. We define a equivalent class of  path $\alpha:I\rightarrow X$ by
\begin{equation}
[I,X]=[I, \big\{1 \big\}]\coprod   [I,\big\{2 \big\}]  \coprod [I, \big\{3 \big\}] \coprod  [I,\big\{3 \big\}] \coprod [I,\big\{3 \big\}] \coprod  [I,\big\{4 \big\}].
\end{equation}
Since $S^{1}$ is homotopy equivalent to $I/\partial I$, we induce a fundamental group of space of time series data,
\begin{equation}
[S^{1},X]=[S^{1}, \big\{1 \big\}]\coprod   [S^{1},\big\{2 \big\}]  \coprod [S^{1}, \big\{3 \big\}] \coprod  [S^{1},\big\{3 \big\}] \coprod [S^{1},\big\{3 \big\}] \coprod  [S^{1},\big\{4 \big\}].
\end{equation}

\begin{figure}[!t]
	\centering
	\includegraphics[width=0.45\textwidth]{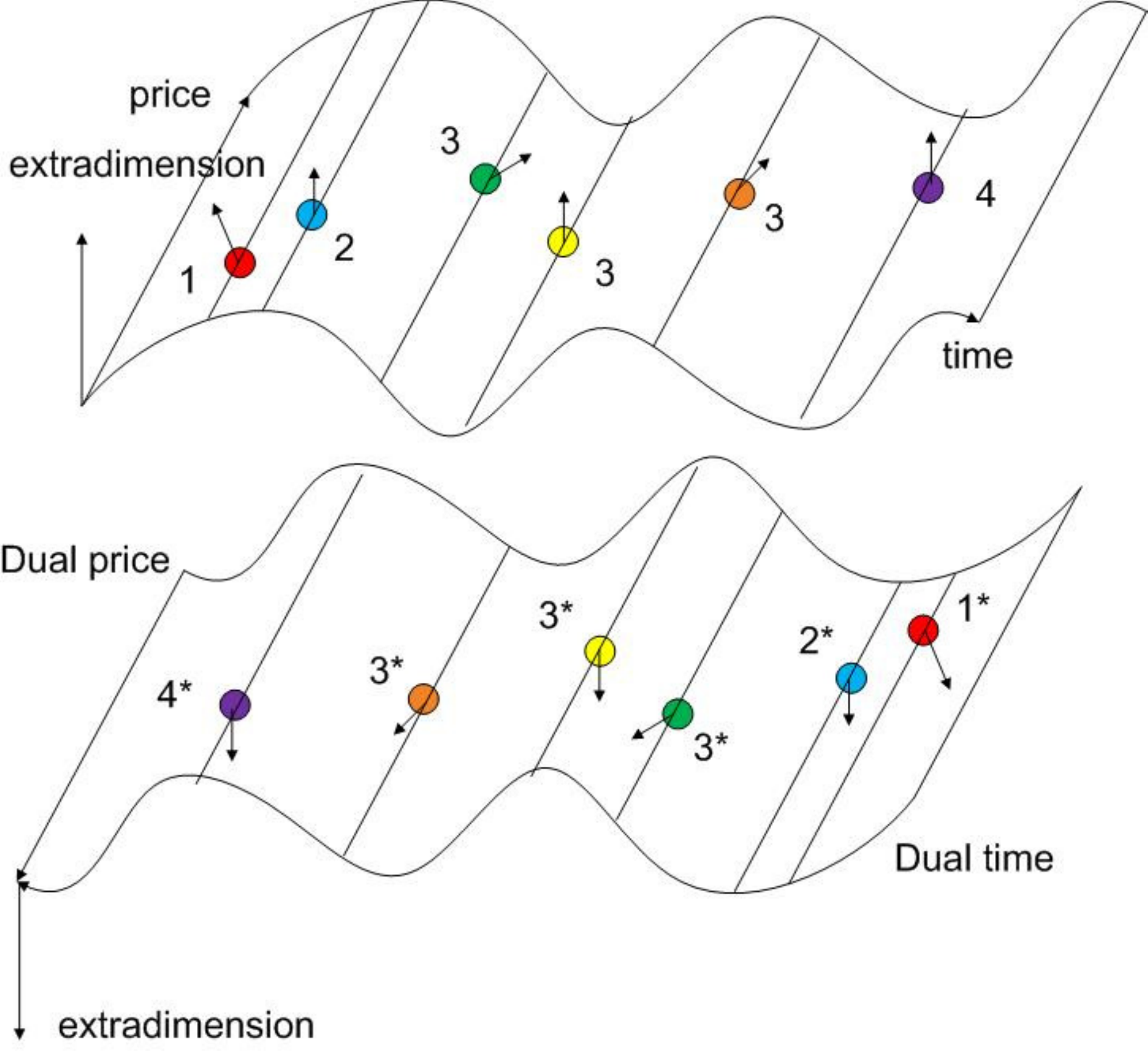}
	\caption{A schematic view of the loop space of time series data with the covering space.\label{loop3}}
\end{figure}

In next section we introduce a precise definition of time series data in loop space $\Omega( X,x_{0})$ (see Fig.~\ref{separate2}). We use equivalent class of loop space in covering space of predefined 4 basis $s_{i},i=1,2,3,4$  in $S^{1}=\{x\in \mathbb{C}, |x|=1\}$ and another perpendicular $S^{1\ast}$ for location of time series data (Fig.~\ref{loop3}). These equivalent classes induce a group structure with a symmetry of time series equivalent to a orbital in quantum state. This is a spinor field of time series data in which one can classify a financial time series data. This new construction used to explain a precise definition of financial time series allows one to search for a new concept of a spinor field of time series data and a mirror symmetry in space of financial time series. In the prove that a space of time series data is a Kolmogorov space we use projective geometry of quaternionic field instead of a discrete topology.

\section{Loop space of time series}\label{sec:loop}

\subsection{Morphology of financial time series}

For a given time series $x_{t}\in X$, we induce two spaces of times series with the empirical mode decomposition method (EMD) \cite{Huang}, a complex time scale in $4$-dimensional space and a physiology of time series in $4$-dimensional vector space $s_{j}\in V\simeq \mathbb{H}$. 

An end point of time series data $\mathrm{End}(x_{t})$ is defined as
\begin{equation}
\mathrm{End}(x_t)= \sum_{i=1}^{4}\lambda_{i} g_{ij}s_{j},
\end{equation}
with $g_{ij}$ -- a Jacobian of transformation of the coordinate system and $\lambda_{i}\in \{0,1\} $. We have
\begin{equation}
\mathrm{End}\big(x_{t}(x*\otimes t* )\big)= 
\lambda_{1}g_{11}\left  [\begin{array}{c}
s_{1}\\
0\\
0\\
0\\
\end{array}\right]
+
\lambda_{2}g_{22}\left  [\begin{array}{c}
0\\
s_{2}\\
0\\
0\\
\end{array}\right]
+\cdots 
\lambda_{4}g_{44}\left  [\begin{array}{c}
0\\
0\\
0\\
s_{4}\\
\end{array}\right].
\end{equation}

\begin{figure}[!t]
	\centering
	\includegraphics[width=0.9\textwidth]{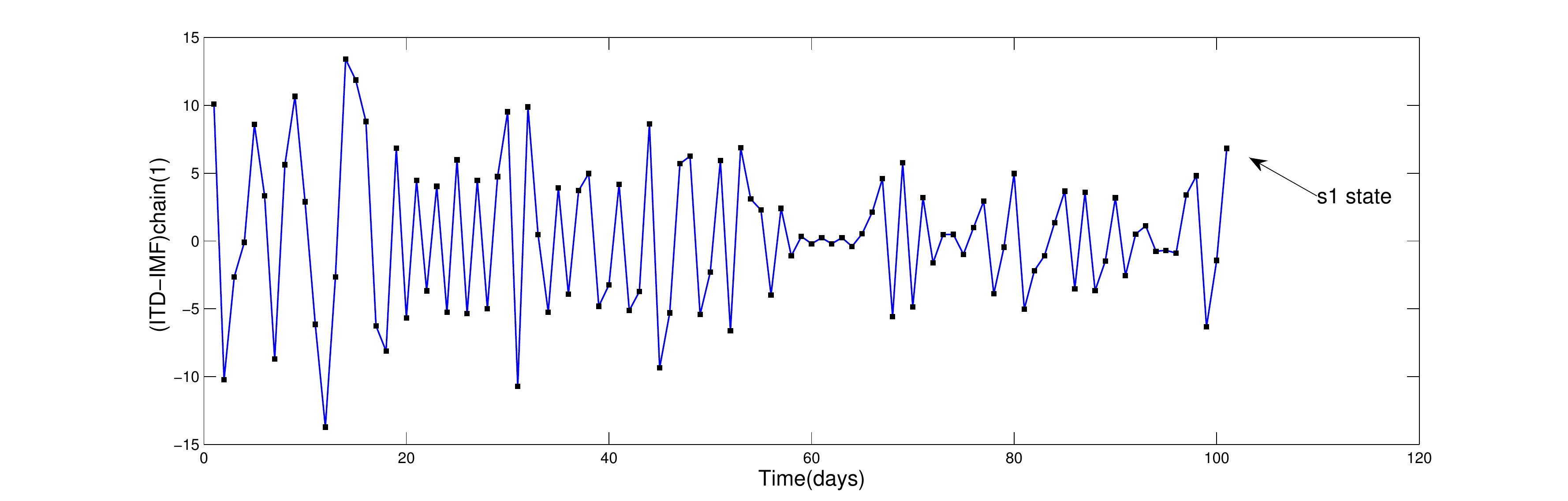}
	\includegraphics[width=0.9\textwidth]{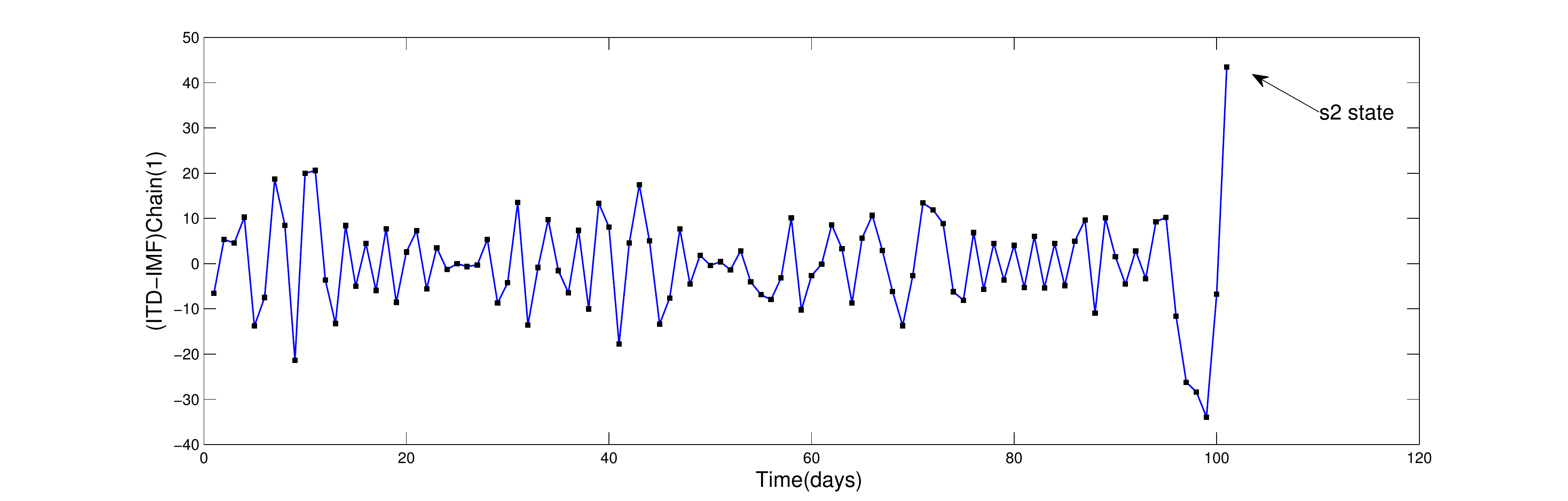}
	\includegraphics[width=0.9\textwidth]{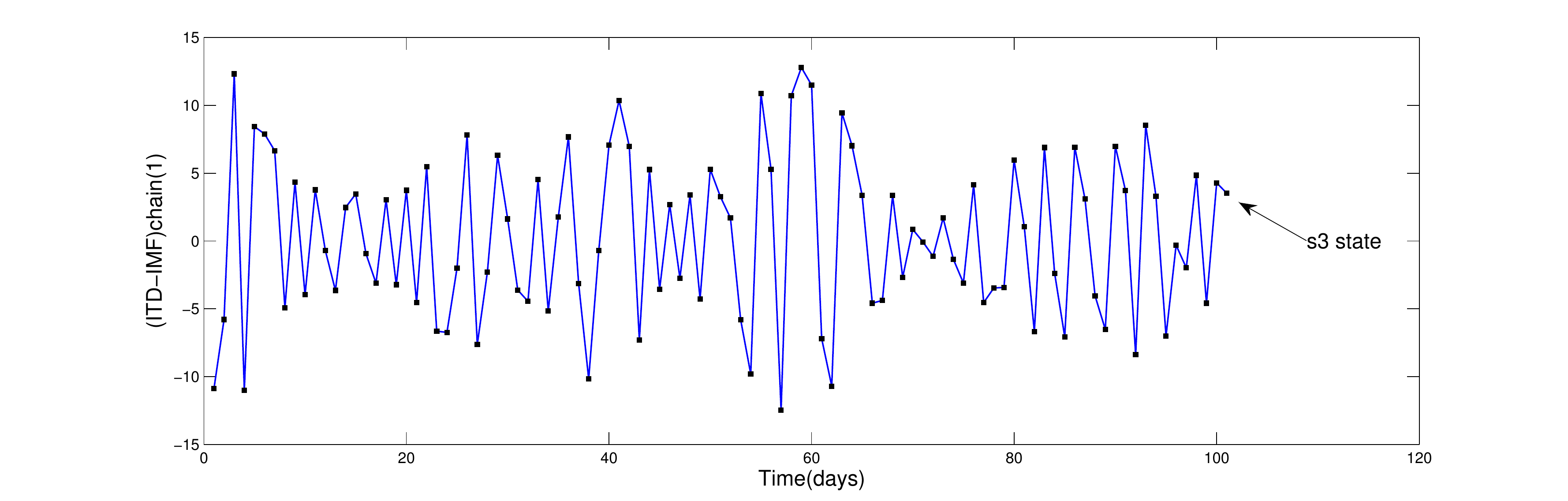}
	\includegraphics[width=0.9\textwidth]{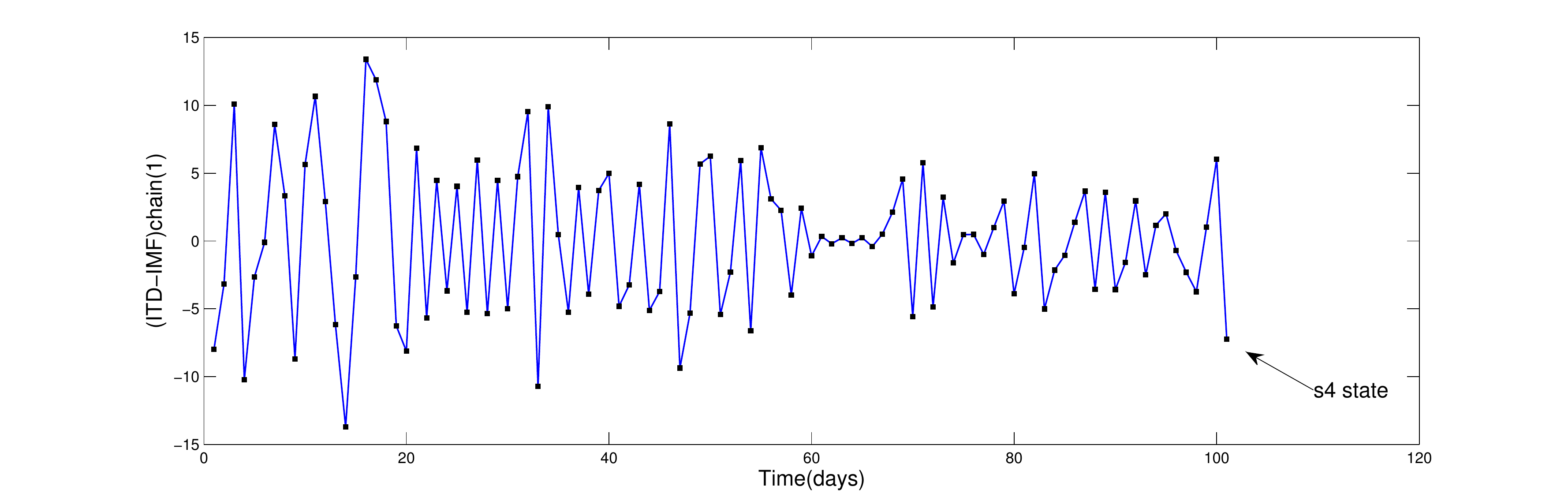}
	\caption{The example of the shape of end point for \ICHAIN\ for $s_1$, $s_2$, $s_3$, $s_4$ states of SET index within 100 days. \label{skeleton}}
\end{figure} 
The hidden direction is coming from a state of the end point for a skeleton of time series data so called $\ICHAIN$. Let four hidden directions of a local state of physiology of end point of time series be
\begin{align} 
s_{1}(x_{t}) &= \mathrm{mono}_{\mathrm{up}}(x_{t}), \nonumber\\
s_{2}(x_{t}) &= \max x_{t}, \nonumber\\
s_{3}(x_{t}) &= \mathrm{mono}_{\mathrm{down}}(x_{t}), \\ \nonumber
s_{4}(x_{t}) &= \min x_{t}. 
\end{align}
where $\mathrm{mono}_{\mathrm{up}}(x_{t})$ is a monotonic function up of time series data of the end point of time series data. It is the point between minimum point of time series data and maximum point of time series data. Sometimes this point does not exist. $\mathrm{mono}_{\mathrm{down}}(x_{t})$ is defined by a monotonic function down of time series data of the end point of time series data. It is the point between maximum point of time series data to a  minimum point of time series data, sometimes this point also does not exist.

\begin{Definition}[Cyclic coordinate of time scale]
	Let $T_{1}$  be a location of monotone function up, measured from a distance between $s_{4}$ and $s_{1}$ of time series data of $\ICHAIN$.  $T_{2}$ be a location of maximum point, measured from distance between $s_{4}$ to $s_{2}$ of time series data of $\ICHAIN$. $T_{3}$ is a location of monotone function down, measured from distance between $s_{4}$ to $s_{3}$ of time series data of $\ICHAIN$. $T_{4}$ is a location of minimum point, measured from distance between $s_{4}$ to next cycle of $s_{4}$ of time series data of $\ICHAIN$. If a restart of every cycle starts from zeros every cyclic time coordinate of $t=(T_{1},T_{2},T_{3},T_{4})$ will be a circle of time scale in loop space.
\end{Definition}
The details of EMD algorithm, the definition of $\ICHAIN$, the skeleton of time series data and the empirical work on financial time series data of cyclic time coordinate are shown in Fig.~\ref{skeleton} and  Appendix~\ref{app:data}.

\subsection{Covering space of time series data}

Let 
\begin{equation}
x_{1}\rightarrow x_{2}\rightarrow x_{3}\rightarrow \cdots \rightarrow x_{n}
\end{equation}
be a time series data with underlying trivial topological based space $X$ where
\begin{equation}
X=\big\{x_1 \big\}\coprod \big\{x_2 \big\}\coprod  \big\{x_3 \big\}\coprod  \cdots \coprod  \big\{x_n \big\}.
\end{equation}
We define a tangent space of time series data by lifting path of covering space. We use a notation

\begin{equation}
\coprod_{i=1}^{n}T_{x_{i}}X=T_{x_1 }X\coprod T_{x_2}X\coprod  T_{x_3}X\coprod  \cdots 
\coprod  T_{ x_n }X.
\end{equation}
for a tangent space of time series data or covering space of time series data. The element of tangent space of time series data is defined 
by 4 states of morphology of time series data denoted by $s_{i},i=1,2,3,4$. We have $dx_{i}\in T_{x_{i}}X$ if $x_{i}\in X$,
\begin{equation}
dx_{i}=\sum_{j=1}^{4}p_{i}\frac{\partial x_{i}}{\partial s_{j}}ds_{j}.
\end{equation}
where $p_{i}$ is a probability to find a hidden state $s_{i}$. Giving a tangent space we also induce a dual tangent space of time series data $T_{x}^{\ast}X$ and also differential form of time series data using wedge product $\wedge T^{\ast}_{x}X$.
\subsection{Tensor field of time series data}

In definition of time series data we assume two independent spaces of two measurement systems, a vector space of price $X$ and a space of time $t\in S^{1}$, respectively a complex unit sphere $S^{1}$. The space of time series data obtained by merging $X$ and $S^{1}$ is represented by the tensor product $X\otimes S^{1}$ as a state space of time series data in tensor field. The states that can be broken into the tensor product of states from the constituent subsystems are called separable states, whereas states that are unbreakable are called entangled states of time series data. Let $x\in X$, $t\in S^{1}$ we have a dual price $x_{\ast}=\vec{t}\wedge \vec{x}$, $x^{\ast}=\vec{x}\wedge\vec{t}$. We can define dual time scale \cite{Dooley} of time series data by
\begin{equation} 
t^{\ast}=x_{\ast}\otimes x^{\ast}.
\end{equation}

\subsection{Homotopy path of time series data}

Let
\begin{equation} 
x_{1}\rightarrow x_{2}\rightarrow x_{3}\rightarrow \cdots \rightarrow x_{n},
\end{equation}
be a sequence of points of financial time series in categories of  SET with preorder relation of time ordering as morphism.

Let $\pi_{1}: \mathrm{TOP} \rightarrow \mathrm{GROUP}$ be a functor of fundamental group of chosen based point $x_{0}$ of object in categories of SET with $(X,x_{0})$ a topological space $X$ of underlying space of financial time series as object in TOP
\begin{equation} 
\pi_{1}(X, x_{1})\rightarrow \pi_{1}(X,x_{2})\rightarrow 
\pi_{1}(X,x_{3})\rightarrow \cdots \rightarrow\pi_{1}(X, x_{n}),
\end{equation}
there is a sequence of equivalent class of loop in loop space of time series data
\begin{equation}\Omega(X, x_{1})\rightarrow \Omega(X,x_{2})\rightarrow 
\Omega(X,x_{3})\rightarrow \cdots \rightarrow\Omega(X, x_{n}).
\end{equation}
There exists a one-to-one sequence of discrete time intervals between measurements 
\begin{equation}
t_{1}\rightarrow t_{2}\rightarrow t_{3}\rightarrow \cdots \rightarrow t_{n-1}
\end{equation}
with a loop space of location (complex time scale coordinate of time series)
\begin{equation}
\Omega(X, t_{1})\rightarrow \Omega(X,t_{2})\rightarrow 
\Omega(X,t_{3})\rightarrow \cdots \rightarrow\Omega(X, t_{n-1}).
\end{equation}

Let $4$-dimensional space $\mathbb{H}$ be a space of cyclic time coordinate. Let
\begin{equation} 
t(x)= T_{1}(x)+ T_{2}(x)i+T_{3}(x)j+ T_{4}(x)k \in \mathbb{H}.  
\end{equation}
where $T_{1}$ is a time from the origin to the state of monotone up $s_{1}$ of time series data of $\ICHAIN$. $T_{2}$ is a time period from the origin to the state of maximum $s_{2}$ of time series data of $\ICHAIN$. $T_{3}$ is a time from the origin to the state of monotone down $s_{3}$ of time series data of $\ICHAIN$. $T_{4} $ is a time period from the origin to the state of minimum $s_{4}$ of time series data of $\ICHAIN$.  $\mathbb{H}$ is a quaternion field with $3$ complex numbers implying hidden states of time scale with $i^{2}=j^{2}=k^{2}=-1$, $ijk=-1$.

For a given  time series data set $X=\{x_{1},x_{2},\cdots, x_{n}\}$ we induce a set of location in time series data $t_{X}=\{t_{1},t_{2},\cdots ,t_{n}\}$, $|X|=|t_{X}|$. Let a functor $[S^{1} \,\cdot]: \mathrm{SET}\rightarrow h-\mathrm{TOP}$, $X\mapsto [S^{1} \, X]=\pi_{1}(X)$  where $S^{1}=\{z\in \mathbb{C}, |z|=1\}.$ 
An object of homotopy category $h-\mathrm{TOP}$ is a set of equivalent class of classifying space $S^{1}$ over a pointed space of time series data. Let a homotopy path be
\begin{equation} 
[\alpha]  \in [S^{1} \, X] 
\end{equation} 
to partition $X$ into $4$ equivalent classes of equivalent location in physiology of time series $[T_{1}]$, $[T_{2}]$, $[T_{3}]$ and $[T_{4}]$.

\begin{figure}[!t]
	\centering
	\includegraphics[width=0.5\textwidth]{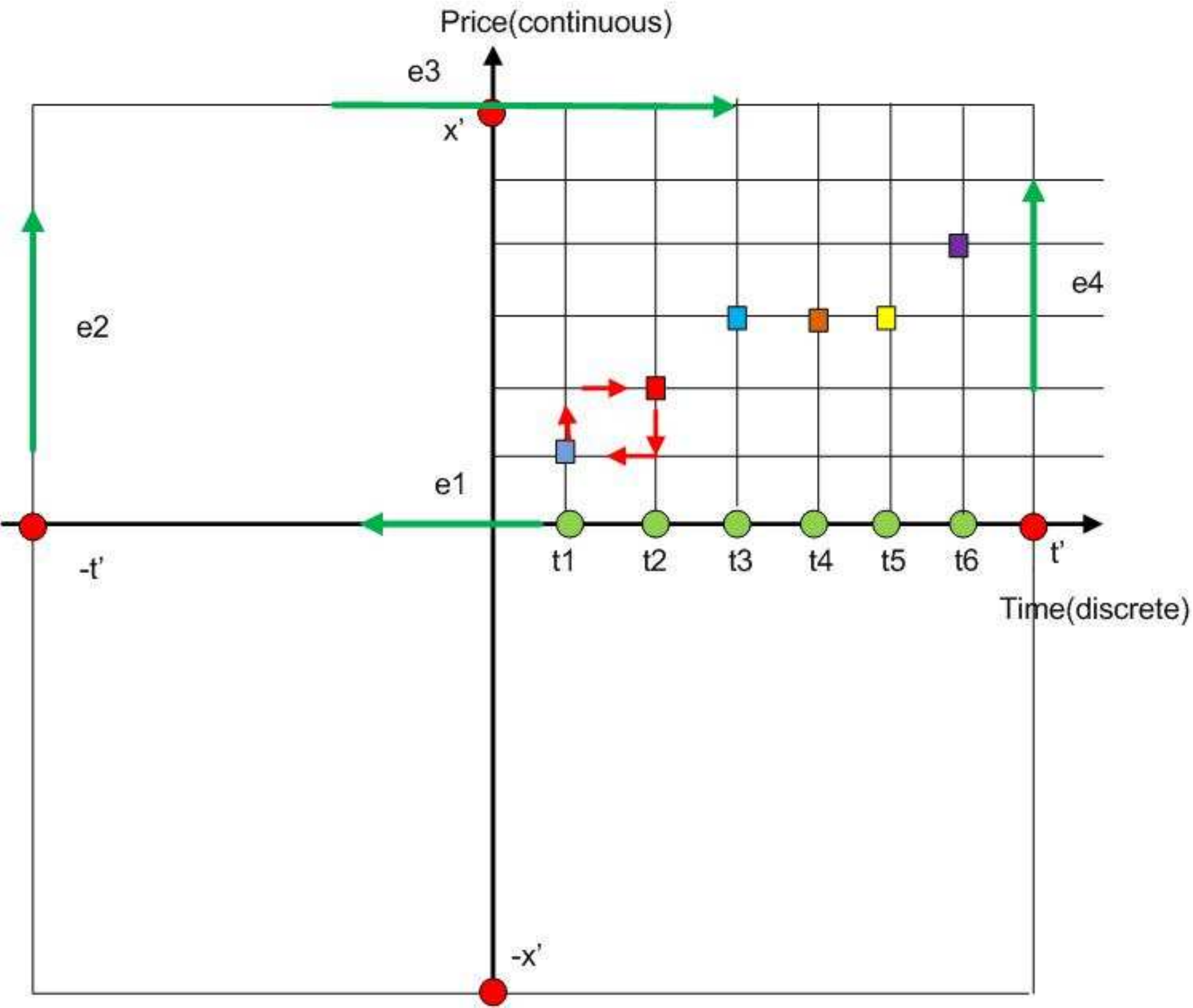}
	\caption{CW complex of boundary of space of time series data. For every value $x_{t}$ of time series data in Euclidean plane there exist an upper bound $x'\in \mathbb{R}$ such that for every $x_{t}\in X_{t},x_{t}<x'$. For index set of time $t_{n}$ there exists $t'>t_{n}$ for every $n\in \mathbb{N}$. Therefore we can define a closed boundary of Euclidean subspace of time series data by rectangle in this figure. After that we define cell decompostion for each corner of boundary in non-orientation state. Then we glue them into Möbius strip to induce spinor field of time series data.\label{loop2}}
\end{figure}

Let us consider correlation problem of inert frame of reference. Most economists use a formula below to calculate a correlation on price not on time scale but on fix time scale $\mathrm{corr}(x_{1},x_{2})=\mathrm{Corr}(x_{1}(t),x_{1}(t))$ with
\begin{equation} 
\mathrm{Corr}(x_{1},x_{2})=\sum \frac{(x_{1}-\mu_{1})^{2}}{\sigma_{1}}\frac{(x_{2}-\mu_{2})^{2}}{\sigma_{2}}   
\end{equation} 
Now we use a transformation to interchange coordinates between price and time (transition path) by inversion (projection on time line and price line) $t'=x(t)$ which induces a bijective map
\begin{equation} 
i:t'(x_{1},x_{2})   \rightarrow x(t_{1},t_{2}).
\end{equation}  
The Jacobian on this coordinate transformation between space and time is so called Minkowski metric. It is opposite to Euclidean space concept by means the space and time are completely separated. We consider the correlation in waiting time $t_{1}$, $t_{2}$ of the join return of rotational invariant and translation invariant of $2$ returns $x_{1}$, $x_{2}$,
\begin{equation} 
\mathrm{Corr}\big(t_{1}(x),t_{2}(x)\big)=\sum \frac{\big(t_{1}(x)-\mu_{1}\big)^{2}}{\sigma_{1}}\frac{\big(t_{2}(x)-\mu_{2}\big)^{2}}{\sigma_{2}}  
\end{equation} 
which induce a Jacobian matrix for the transformation between these two ways to calculate a correlation by using price and time coordinate,
\begin{equation} 
J= \left |\begin{array}{cc}
\frac{\partial t_{1}}{\partial x_{1}} &\frac{\partial t_{1}}{\partial x_{2}}\\
\frac{\partial t_{2}}{\partial x_{1}} &\frac{\partial t_{2}}{\partial x_{2}}\\
\end{array}\right |
\end{equation} 
where $x=x(x_{1},x_{2})$, $dr=J(dt)$. In Euclidean space-time continuum concept it is impossible to use this kind of transformation. What we can do is to transform both space (price) and time all together because price and time of two stocks in which they are correlated to each other are evolved together and cannot be separated (isometry) so we have Jacobian of transformation as a part of isometry group 
\begin{equation} 
G\langle x,t\rangle=\langle x,t\rangle,\quad G=J^{2}
\end{equation} 
so we have
\begin{equation} 
d(x',t')= J( d(x,t))
\end{equation} 
with $J=-1$ to interchange a projection between price and time as inversion point of symmetry breaking. We define this induced Jacobian transformation in hidden coordinate of complex plane by introducing additional hidden coordinate of a projection to imaginary axis of complex plane by
\begin{equation} 
J= \left |\begin{array}{cc}
x &t\\
x*&t*\\
\end{array}\right |=x\wedge t* -x*\wedge t.
\end{equation} 
The determinant commute by using wedge product with hidden coordinate defined by induce cross product of vector of space and time and hidden space and time as double complex plane. The empirical work of correlation between price and time is given by correlation matrix between cyclic time coordinate $t=(T_{1}$, $T_{2}$, $T_{3}$, $T_{4})$ and cyclic state $x=(s_{1}$, $s_{2}$, $s_{3}$, $s_{4})$,
\begin{equation} 
\mathrm{Corr}(x,t)= \left [\begin{array}{ccccc}
&T_{1} &T_{2}&T_{3}&T_{4}\\
s_{1}& \mathrm{Corr}(s_{1},T_{1}) &  \mathrm{Corr}(s_{1},T_{2})& \mathrm{Corr}(s_{1},T_{3}) & \mathrm{Corr}(s_{1},T_{4})\\
s_{2}&  \mathrm{Corr}(s_{2},T_{1}) &  \mathrm{Corr}(s_{2},T_{2})& \mathrm{Corr}(s_{2},T_{3}) & \mathrm{Corr}(s_{2},T_{4})\\
s_{3}&   \mathrm{Corr}(s_{3},T_{1}) &  \mathrm{Corr}(s_{3},T_{2})& \mathrm{Corr}(s_{3},T_{3}) & \mathrm{Corr}(s_{3},T_{4})\\
s_{4}&   \mathrm{Corr}(s_{4},T_{1}) &  \mathrm{Corr}(s_{4},T_{2})& \mathrm{Corr}(s_{4},T_{3}) & \mathrm{Corr}(s_{4},T_{4})\\
\end{array}\right ].
\end{equation} 
and is shown in Appendix~\ref{app:data}. We define each point of $T_{i}$, $i=1,2,3,4$ as disjoint pointed space embedded in complex projective space as cell decomposition of pointed space $e_{0}$ of CW complex decomposition. We can define a point of complex time scale as choosen basepoint for base space of fundamental group. When we collapse a Riemann sphere to base point of time scale of time series we get a cone space of time series as quotient topology. We have $e^{0}=\{T_{i}\}$, $T_{i}\in S^{0}\subset S^{1}\subset S^{2}\subset S^{3}\subset\cdots$. CW-complex (Fig.~\ref{loop2}) is required only 2 cells, one a point $e_{0}$ and the other $e^{n}=s^{n}-\{e^{0}\}$ which is homeomorph of ball $\mathbb{B}^{n}$. We define our translated Riemann sphere with relative coordinate of time series by attaching cell $e^{0}=\{T_{i}\}$ to a center of Riemann sphere as relative frame of time scale of Riemann sphere $S^{2}$.

\begin{figure}[!t]
	\centering
	\includegraphics[width=0.55\textwidth]{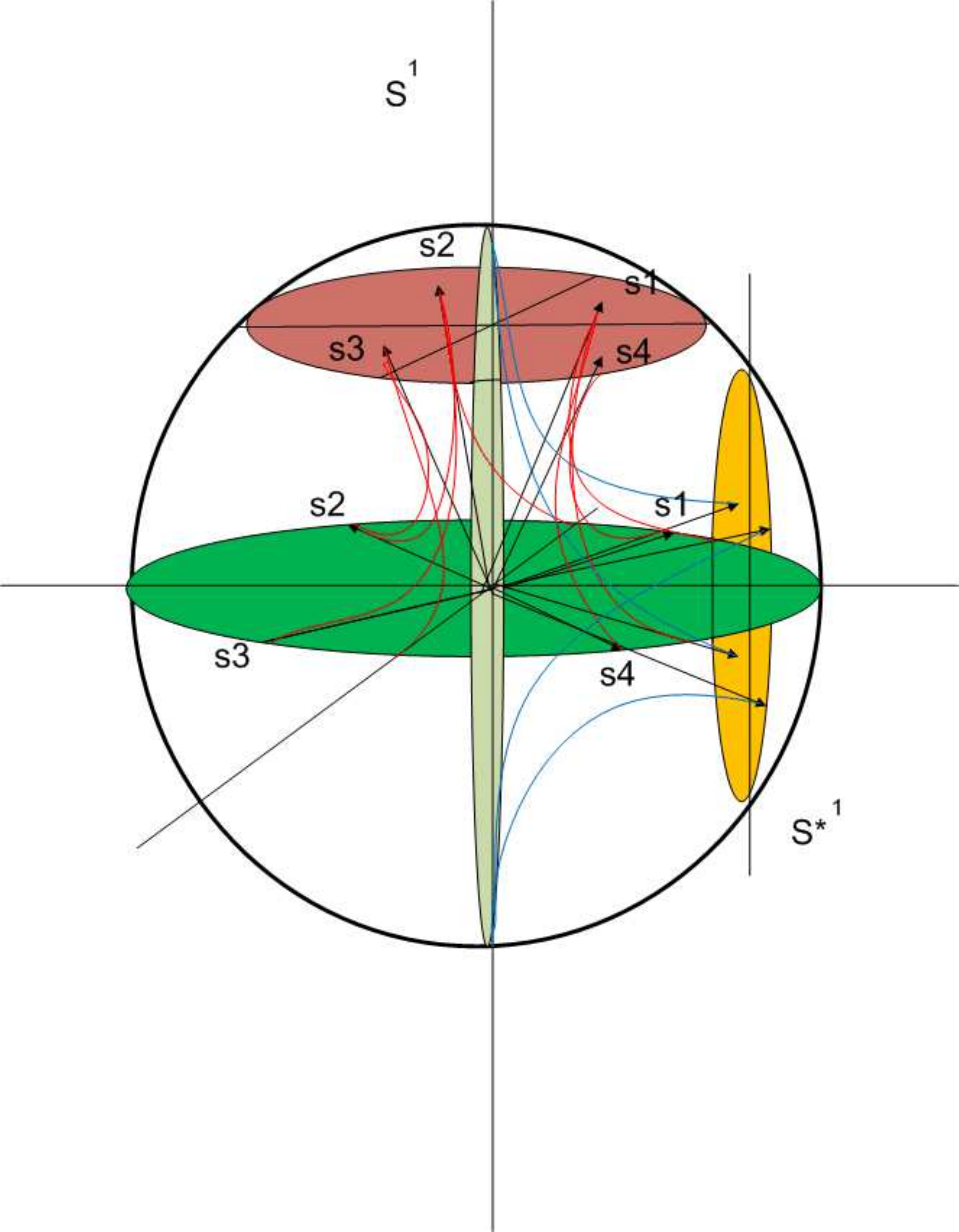}
	\caption{A homotopy path of hyperbolic space of time series. One can see that hyperbolic space of time series data is inside Riemann sphere of time series data. One side of sphere is predictor state of time series data,  the perpendicular side is predictant state of time series data.	The hyperbolic line connects the predictor and predictant states.\label{collapse22}}
\end{figure}

If we take into account only equivalent class of loop  in $\pi_{1}(X,x_{t})$, a loop structure of time series data in empirically measured by Hilbert transform of $\ICHAIN$. The result of Hilbert transform of time series data is a cycle in complex plane in which it is homotopy equivalent to $S^{1}$.  We can explicitly define a physiology of time series data in loop space by using equivalent class of path in $S^{1}$. We separate $S^{1}$ into $4$ states of physiology of time series data in following way. 

Let $s_{1}\in [ s_{1}(x_{t})]$ be an equivalent class of loop of time series from $x_{0}$ to $s_{1}(x_{0})=e^{0\ii}\in S^{1}$ and $s_{1}(x_{1})=e^{\ii\frac{\pi}{2}}\in S^{1}$, a covering space of time series  with homotopy path (see Fig.~\ref{collapse22})
\begin{equation} 
h: S^{1}\times [0,T_{1}] \rightarrow   S^{1},\qquad h(t,0)=e^{0\ii},\quad h(t ,T_{1})=e^{\ii\frac{\pi}{2}}.
\end{equation} 

Let $s_{2}\in [s_{2}(x_{t})]$ be an equivalent class of loop of time series from $x_{0}$ to $s_{2}(x_{0})=e^{\ii\frac{\pi}{2}}\in S^{1}$ and  $s_{2}(x_{1})=e^{\ii\pi}\in S^{1}$, a covering space of time series with homotopy path
\begin{equation} 
h: S^{1}\times [T_{1},T_{2}] \rightarrow   S^{1},\qquad h(t,T_{1})=e^{\ii\frac{\pi}{2}},\quad h(t ,T_{2})=e^{\ii\pi}.
\end{equation} 

Let $s_{3}\in [s_{3}(x_{t})]$ be an equivalent class of loop of time series from $x_{0}$ to $s_{3}(x_{0})=e^{\ii\pi}\in S^{1}$ and $s_{3}(x_{1})=e^{\ii\frac{3\pi}{2}}\in S^{1}$, a covering space of time series with homotopy path
\begin{equation} 
h: S^{1}\times [T_{2},T_{3}] \rightarrow   S^{1},\qquad h(t,T_{2})=e^{\ii\pi},\quad h(t ,T_{3})=e^{\ii\frac{3\pi}{2}}.
\end{equation} 

Let $s_{4} \in [s_{4}(x_{t})]$ be an equivalent class of loop of time series from $x_{0}$ to $s_{4}(x_{0})=e^{\ii\frac{3\pi}{2}}\in S^{1}$ and $s_{4}(x_{1})=e^{\ii2\pi}\in S^{1}$, a covering space of time series with homotopy path
\begin{equation} 
h: S^{1}\times [T_{3},T_{4}] \rightarrow   S^{1},\qquad h(t,T_{3})=e^{\ii\frac{3\pi}{2}},\quad h(t,T_{4})=e^{\ii2\pi}.
\end{equation} 

\begin{figure}[!t]
	\centering
	\includegraphics[width=0.8\textwidth]{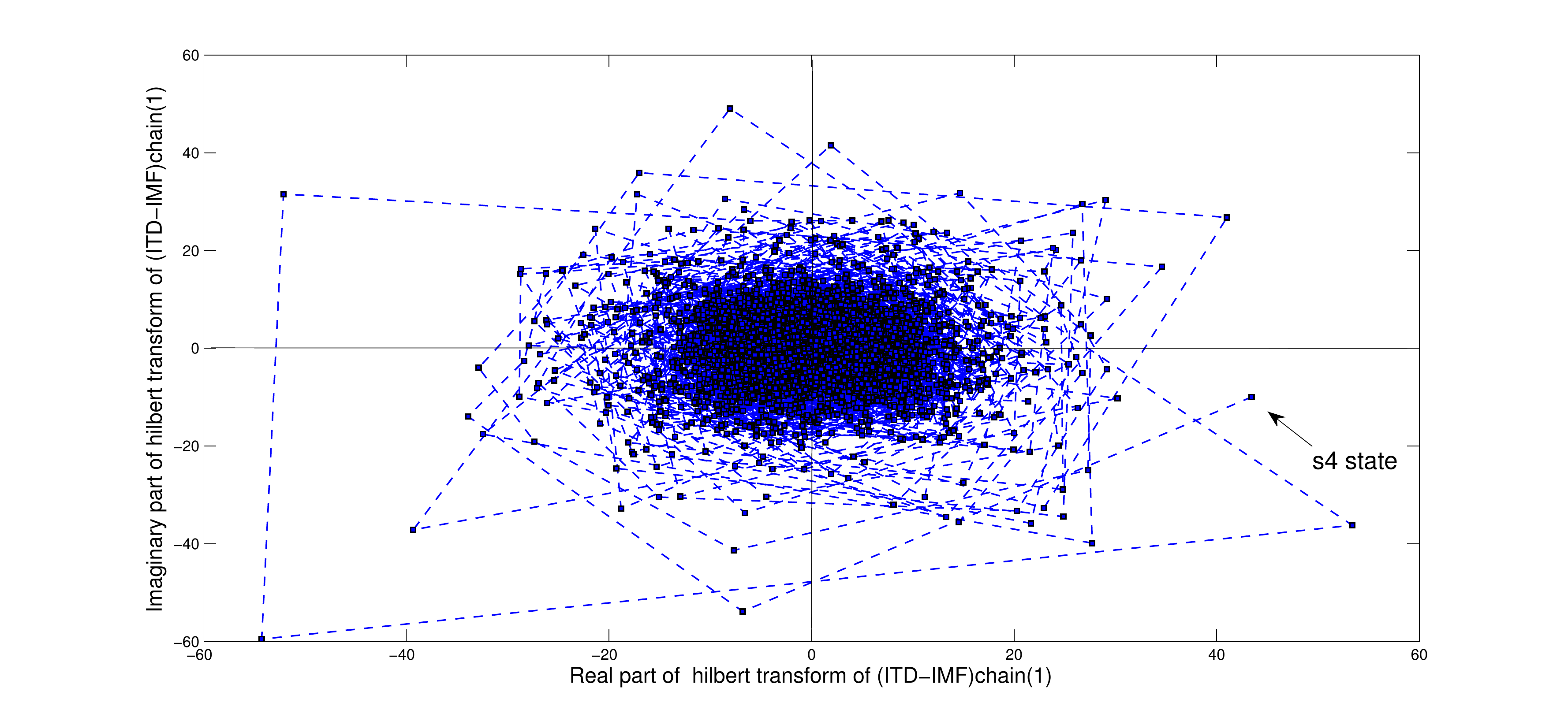}
	\includegraphics[width=0.45\textwidth]{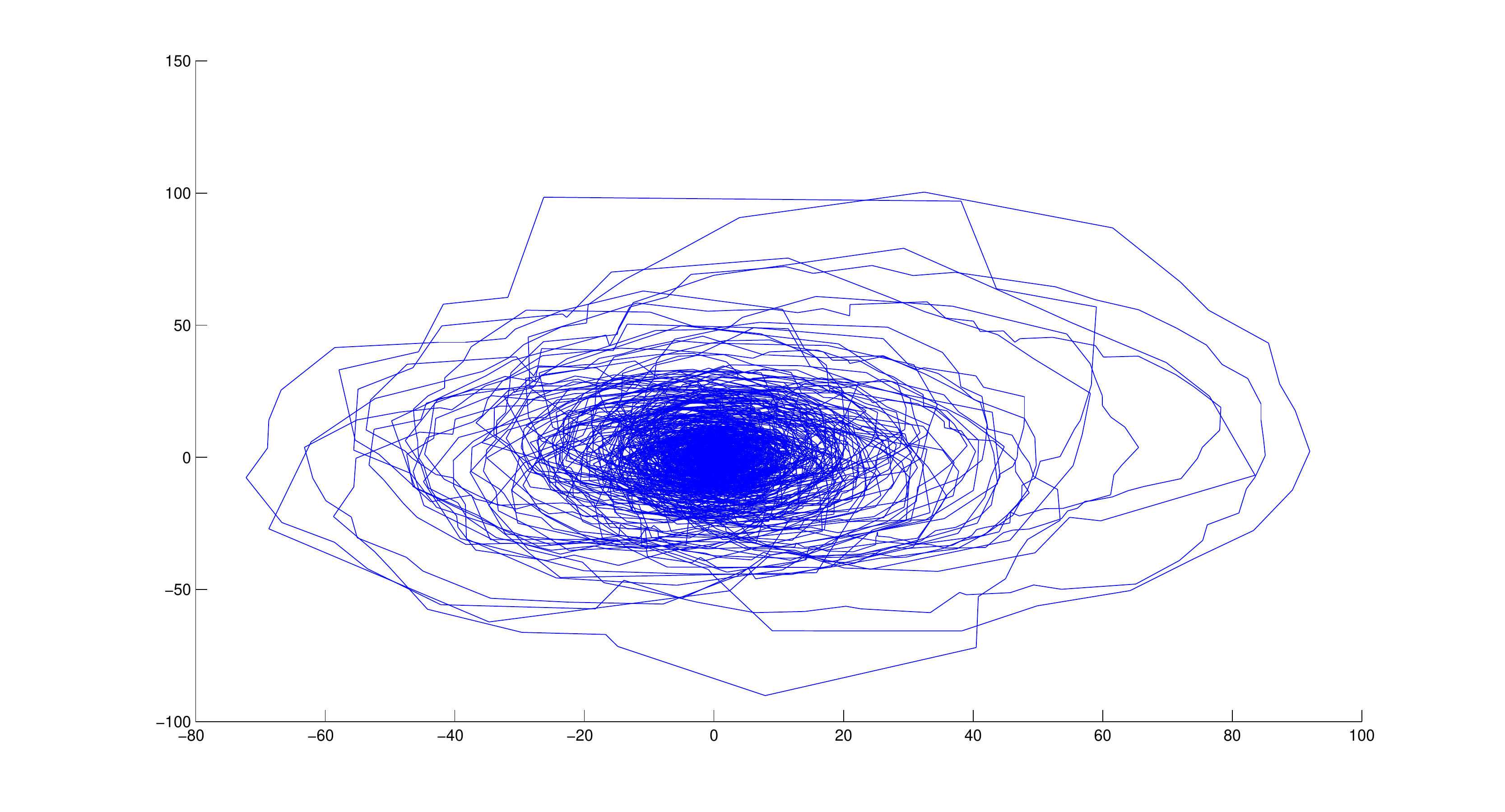}
	\includegraphics[width=0.45\textwidth]{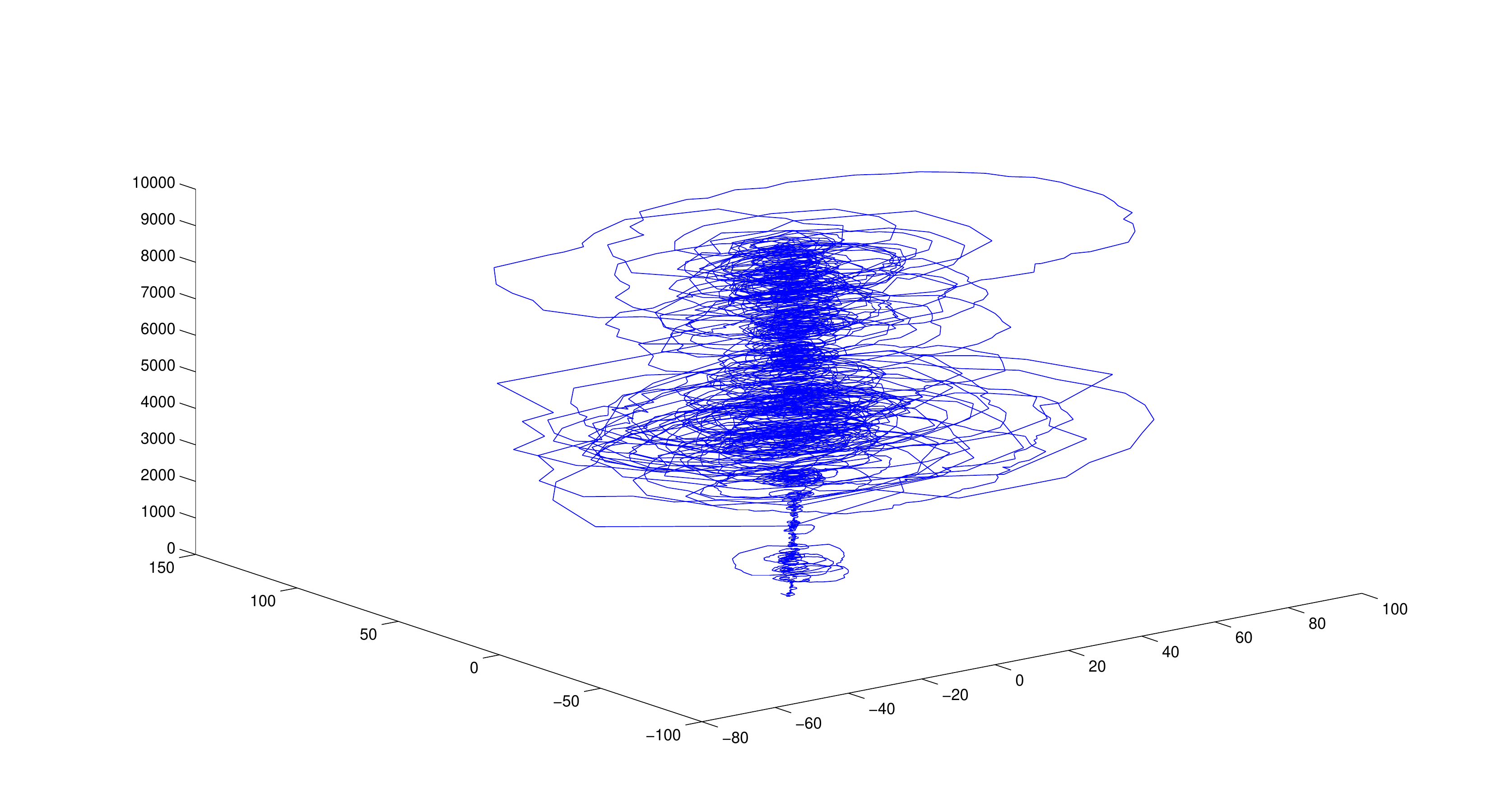}
	\caption{The upper figure shows the graph of Hilbert transformation of $\ICHAIN$ of SET index of $100$ daily closed price with the	end point of time series in $s_{4}(x_{t})$ state. The bottom figure on the left represents Hilbert transformation of $\ICHAIN$ of SET index of $1000$.	The bottom right figure represents the same plot in $3$-dimensional view. \label{hilbert2}}
\end{figure}

\begin{figure}[!t]
	\centering
	\includegraphics[width=0.45\textwidth]{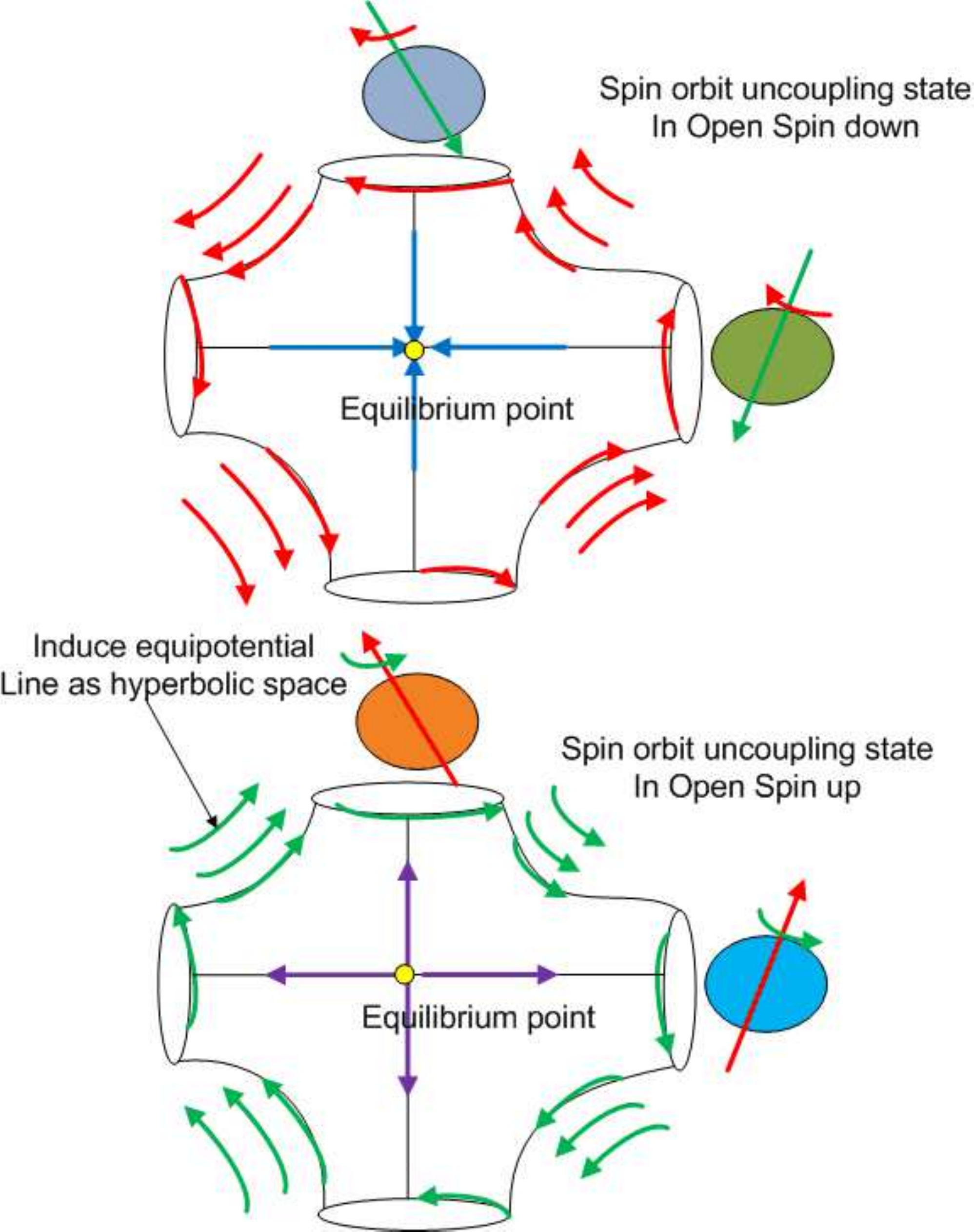}	
	\caption{A spin orbit coupling state between predictor and predictant states of spinor field of time series data induces an equipotential line as hyperbolic state for time series data.	A path of expectation of physiology of time series is defined by this equipotential line of evolution feedback path between predictor and predictant. Uncoupling spinor fields of time series data can induce two types of fix pointed fields in Kolmogorov space for time series data. There exist $8$ states of induced spinor field of time series data represented by red and green hyperbolic equipotential lines.\label{expectpath}}
\end{figure}

An example of empirical analysis of Hilbert transform of $s_{4}$ state of $\ICHAIN$ of financial data is shown in Fig.~\ref{hilbert2}.

An identification of equivalent class coming from an opposite direction of monotone function up and down and also maximum state and minimum state
\begin{equation*} 
[s_{1}]=-[s_{3}],
\end{equation*} 
thus the inverse of $s_{1}$ is $s_{3}$. Since
\begin{equation*} 
[s_{2}]=-[s_{4}]
\end{equation*} 
we have an inverse of $s_{2}$ to be $s_{4}$. We have
\begin{gather*}
[s_{1}]+[s_{3}]=[0]\\
[s_{2}]+[s_{4}]=[0]
\end{gather*}
therefore
\begin{equation} 
[s_{1}]+[s_{3}]+[s_{2}]+[s_{4}]=[0]
\end{equation} 
with $[0]$ be a loop from the origin of time series in ground state to itself.

Therefore, in these constructions we allow all mixed states between $s_{i}$ and expected path in hyperbolic space to $s_{i}^{\ast}$ in complex structure of spinor field of $2$ perpendicular cycles inducing a hidden field between each other as shown in Fig.~\ref{collapse22}. The suitable mathematical model can use a  torus of time series data instead of Riemann sphere. The equipotential line of induced field between evolutional feedback path between predictor state $[s^{\ast}_{i}]$ and predictant state (a real state) is shown in Fig.~\ref{expectpath}.

\subsection{Quaternionic projective space}

In this section we explain the origin of a spinor field of time series data. When we consider time series data in Euclidean plane we have a upper bound of value of measurement in real line $x'>x_{t}$ for all $t$. In time coordinate we have a partial ordering of time scale so we have also an upper bound in time variable $t'>t_{i}$ for all $i$. We will induce $-t'$, then we can connect $t'$ with $-t'$ and define cell decomposition as shown in  Fig.~\ref{loop2}. We glue $e_{1}\sim e_{3}$ and $e_{2}\sim e_{4}$. Then we get a spinor field of time series data as Möbius trip of space of time series data. 

Let $D=\{1,i,j,k\}$ be the canonical basis for set of location $\{T_{1},T_{2},T_{3},T_{4}\}$ in $\mathbb{R}^{4}$. A real quaternion for time series is
\begin{equation}
x(t) = s_{1}(t)+s_{2}(t)i+s_{3}(t)j +s_{4}(t)k\in  \mathbb{H}_{x(t)},
\end{equation}
this coordinate is a cyclic coordinate of value for time series.

Let $\{s_{1},s_{2},s_{3},s_{4}\}$ be a set of value in $\mathbb{R}^{4}$. A real quaternion for time series is
\begin{equation}
t(x)=  T_{1}(x)+T_{2}(x)i+T_{3}(x)j +T_{4}(x)k  \in  \mathbb{H}_{t(x)}.
\end{equation}

Let a mathematical definition of time series be a map between two quaternionic fields to quaternionic projective space $\mathbb{H}_{x_{t}} \times \mathbb{H}_{t(x)} \rightarrow       \mathbb{H}P^{1}\simeq S^{7}/\mathrm{Spin}(3)$, where $\mathrm{Spin}(3)$ is a fibre state of time series with spinor field invariant property. It is an equivalent class of time series of glueing a state $T_{4}$ with $T_{1}$ state for a next cycle of cyclic coordinate.

For a given sequence $x(t)\in \mathbb{R}$ it is known that $\mathbb{H}P^{1}/\mathrm{Spin}(3)=S^{7}$. Let $S^{7}=\{\varphi=(x_{1},x_{2},\cdots x_{8})\in \mathbb{R}^{8}, |\varphi|=1 \}$ be a hidden dimension of financial time series. In the next section we are going to prove that a space of time series in canonical form as defined above is a Kolmogorov space.

\begin{figure}[!t]
	\centering
	\includegraphics[width=0.45\textwidth]{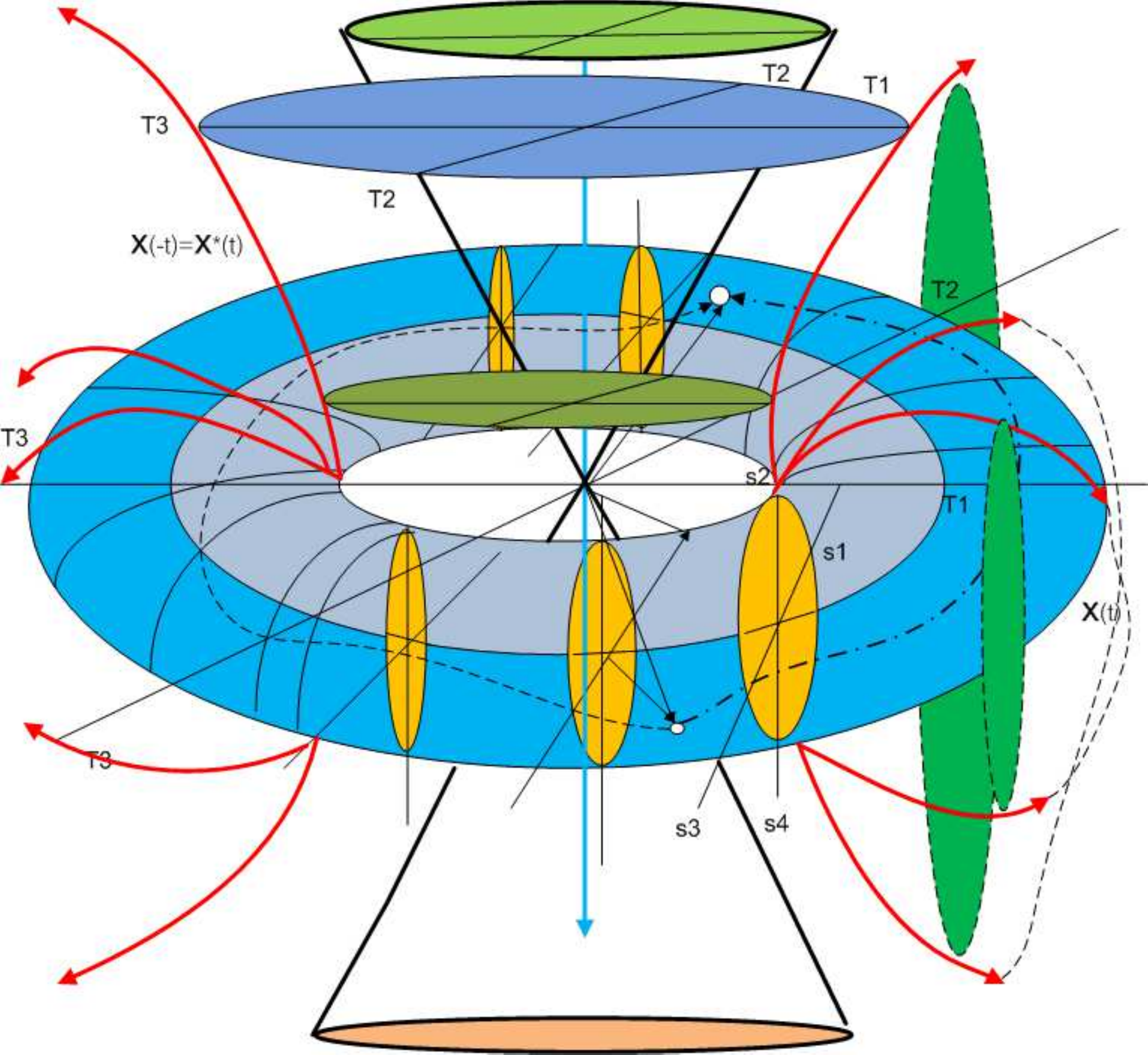}
	\caption{A time series model of  Kolmogorov space. The cyclic coordinate of time is perpendicular to physiology coordinate $s_{1},s_{2},s_{3},s_{4}$ of time series data. All points are separated and we prove that the complex projective plane of time series data in $\mathbb{R}^{8}\cup \{\infty\} =S^{7}$ is disjoint open set of separated point in  $\mathbb{C}P^{3}$.	In this space a time series data can induce an equipotential line a long hidden dimensions. \label{cw}}
\end{figure}

\section{Proof of the main theorem}\label{sec:proof}

We knew that the generalization of Euclidean space of time series in $\mathbb{R}^{n}$ is $n$-dimensional manifold. A local coordinate is defined  as section of tangent of manifold together with Jacobian of coordinate transformation appeared as cocycle of group action over fibre bundle of tangent of manifold in which it is diffeomorphic to $\mathbb{R}^{n}$, when we assume that $n$ data of time series data is embedded in $n$-dimensional manifold $X$, as underlying hidden topological space with value in tangent of manifold $x_{0}\in T_{x_{0}}X=p^{-1}(X)$. In this case we will induce a sequence of tangent bundle of manifold of time series data in pointed space $(X,x_{0})$, with disjoint union of covering space of time series data of choosen based points satisfying with sequence of time series data of $\{x_{0},x_{1},\cdots,x_{n}\}$
\begin{equation}
\begin{CD}
E=\sqcup_{i=0}^{n}T_{x_{i}}X @>>> \{x_{i}\}\in U\times\mathbb{R}^{n}\\
@VVpV             @VVV\\
x_{i}\in U\subset X   @>>>          U\subset X
\end{CD},  
\end{equation} 
where $E$ is a covering space of time  series, $X$ is a $n$-dimensional manifold of time series data with open set $U_{i}$, $F$ is a fibre space $\mathbb{R}^{n}$ of time series data. Most people assume that a sequence of measurements is independent of period and substitute a discrete fibre from $\mathbb{Z}$ with $\mathbb{R}^{n}$. We knew that a tangent space of $\mathbb{R}^{n}$ is $T\mathbb{R}^{n}=\mathbb{R}^{n}$ so we use covering space of $X=\mathbb{R}^{n}$ for time series data in this case. 

Let us assume that a time series data is embedded in non-Euclidean plane, a high dimensional sphere $S^{n}$ with $2$-extradimensions of $S^{1}$ for fibre and hidden fibre induces hidden state $t^{\ast}\in S^{1}$ and $x^{\ast}\in S^{1}$ (see Fig.~\ref{cw}). In this case we model a time series in Riemann sphere $S^{2}=\mathbb{C}P^{1}$ as based space with fibre in $S^{1}$  and covering space in $4$-dimensional space $S^{3}$. This is a principle bundle with discrete fibre $S^{1}$. This construction of principle $\mathrm{Spin}(3)$ bundle (spinor field of time series data) also allows us to define a loop structure between all data of time series connected to each other as path components. In this paper a time series model can be defined by a covering space of $\mathbb{H}P^{1}$ (Fig.~\ref{cw}).

\subsection{Proof of the main theorem}

\begin{Theorem}
	A space of financial time series data is a covering space $S^{7}$ with based space in $X\simeq \mathbb{H}P^{1}$.
	It is a Kolmogorov space with $T_{0}$-separation axiom.
\end{Theorem}

\emph{Proof:}

$\mathbb{H}P^{1}$ is CW complex with one cell for each dimension $k\leq 1$. A cell complex is a Housdorff space satisfying the separation axiom $T_{2}$  which imply Kolmogorov space with lower $T_{0}$-separation axiom. We triangulate $\mathbb{H}P^{1}$ into the union of disjoint subsapce $\{e_{\alpha}, \alpha\in \Lambda \}$ called cells with $e^{n}=S^{n}-\{(1,0,0,\cdots ) \}\subset \mathbb{R}^{n+1}$ with $n$-skeleton space of $X$
\begin{equation}
X^{n}=\cup _{k\leq n}e_{\alpha}^{k}.
\end{equation}

Let $e_{\alpha}$ be an $n$-cell, then there exist a characteristic map of pairs
\begin{equation}
\chi_{\alpha} :(\mathbb{B}^{n},S^{n-1})   \rightarrow (X,X^{n-1})
\end{equation}
which restricts to $\mathbb{B}^{n}-S^{n-1}$ as homeomorphism onto $e_{\alpha}$.

Let $q:S^7\to  \mathbb{H}P^1$ be the quotient map of principle $\mathrm{Spin}(3)$ bundle, and let $u,v\in \mathbb{H}P^1$ with $u\ne v$, there are $x,y\in S^7$ such that $q^{-1}[\{u\}]=\{x,-x\}$ and $q^{-1}[\{v\}]=\{y,-y\}$. Let $\epsilon=\frac13\min\{\|x-y\|,\|x+y\|\}$ and set $U=B(x,\epsilon)\cap S^7$ and $V=B(y,\epsilon)\cap S^7\;$, where the open balls are taken in $\mathbb R^8$. Then $U$, $V$, $-U$ and $-V$ are pairwise disjoint open neighborhoods of $x$, $y$, $-x$ and $-y$, respectively, in $S^7$. Moreover, $q^{-1}\big[q[U]\big]=-U\cup U$ and $q^{-1}\big[q[V]\big]=-V\cup V$. Therefore $q[U]$ and $q[V]$ are disjoint open neighborhoods of $u$ and $v$ in $\mathbb{H}P^1$.

\section{Discussion and conclusion}\label{sec:concl}

\subsection{Discussion}
Recently, there appeared the empirical work \cite{Stanley} which is using the concept of distance between states of financial time series data. The authors classify market state into $8$ states  based on the similarity measure under presumption of separation axiom of topological space. There exists many questions on how $2$ states in financial market can be separated under space of time series data and why stock market has $8$ dimensions as underlying space of financial time series data. There could exist a hidden behavior of spinor field of underlying $8$-dimensional space of time series data with spin invariant property justifying empirical work of $8$ states in financial time series data in quantum mechanics approach since the data can be plot in other higher dimensional coordinate system with taking into account a spinor field. When we project the financial data in real line, we can not take into account the spin behavior of financial time series data. In most cases, $8$-dimensional $S^{7}$ covering space of underlying space of quaternionic field is a most potential candidate space for underlying space of time series as Kolmogorov space with an equation of underlying time series data as spinor field. Many scientists from signal processing \cite{Pei} start to use quaternions for time series data analysis. In physics, there exists a tensor field on Hilbert space for processing information, so called qubit states of quantum information \cite{Planat}.

From the work on quantum entanglement states \cite{Mosseri} under Hopf fibration \cite{Hopf, Lyon} on $S^{7}$ it is known that there exists a $8$-dimensional space of quantum information data. We can borrow the definition with spinor property to redefine a space of financial time series data and explain a situation of eight hidden states in financial market model with spinor field.

\subsection{Conclusion}
We found an existence of spinor field in Kolmogorov space for time series data with $8$ hidden equilibrium states in $8$ hidden dimensions $\mathbb{R}^{8}$ in principle  bundle of physiology of time series data by using Hopf fibration in $S^{7}$.

We proved that a space of time series over cyclic coordinate of location to local  maximum and local minimum state of time series is a Kolmogorov space. We used lifting path of covering space of open set of equivalent class of loop space of time series with open set for separation in unit cycle imply buy and sell operation of trader in cyclic order as quotient group of principle bundle. All quotient topology is a Kolmogorov space. Therefore we proved that the principle bundle of time series is a Kolmogorov space. We hope this approach will be useful for the study of financial time series data.

To show the possibilities of our concept, the pilot empirical data analysis with cyclic coordinate in financial time series data by using empirical mode decomposition and intrinsic time scale decomposition were performed. The preliminary results are presented in the Appendix~\ref{app:data}.

\section*{Acknowledgment}
\addcontentsline{toc}{section}{Acknowledgment}

The authors would like to thank for Chulalongkorn University 100 years scholarship for support this research fund. The work was partly supported by VEGA Grant No. 2/0037/13. R. Pincak would like to thank the
 TH division at CERN for hospitality.

\bibliographystyle{IEEEtran}
\bibliography{kolmogorov}

\newpage
\appendices

\section{Data Analysis}\label{app:data}

\subsection*{Empirical mode decompostion and intrinsic time scale decompostion of time series data}

\begin{figure}[!t]
	\centering
	\includegraphics[width=0.6\textwidth]{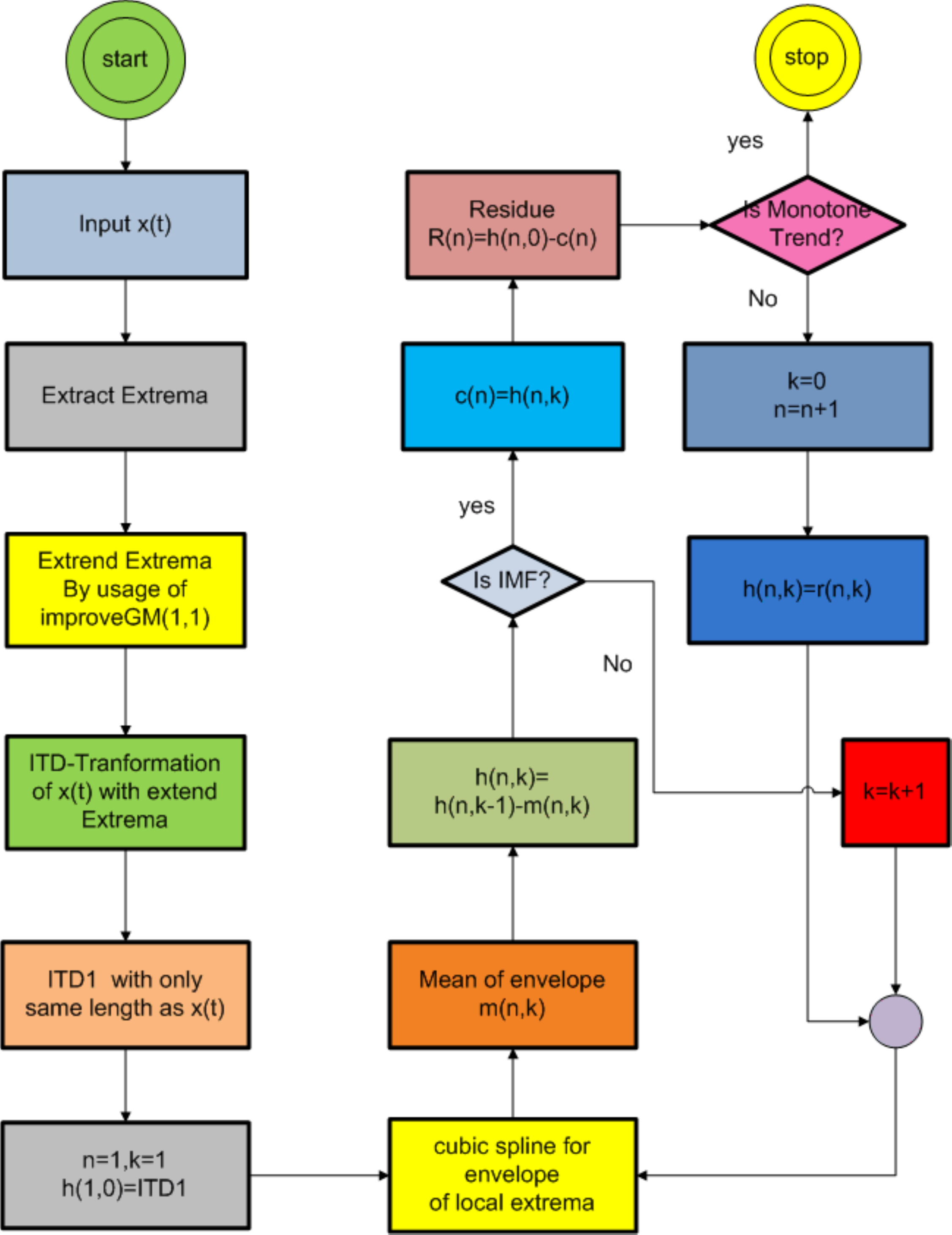}
	\caption{The flowchart of EMD algorithm with endeffect solving whith improving $\mathrm{GM}(1,1)$ and ITD. The algorithm predicts entanglement state of loopback between maximum and minimum statse in time series. \label{algorithm}}
\end{figure}

The algorithm of a shifting process with endeffect solving with improving $\mathrm{GM}(1,1)$ and ITD before sending to empirical mode decompostion (EMD) is shown in Fig.~\ref{algorithm}.

\subsection*{Ensemble empirical mode decomposition}

In ensemble EMD (EEMD) a white Gaussian noise (WGN) is added directly to the signal of interest before applying EMD \cite{EEMD}. The perturbed signal is given by
\begin{equation} 
x_{v}(t)=x_{t}+v_{t},\label{eemd}
\end{equation}
where $x_{t}$ is the input time series and $v_{t}$ is the standard deviation of noise. For a given time series $x_{t}$, the stepwise procedure of EEMD algorithm can be summarized as follows
\begin{description}
	\item[Step 1.] Perturb the input signal $x_{t}$ as described by Eq.~(\ref{eemd}).
	\item[Step 2.] Apply the EMD algorithm to $x_{v}(t)$ to obtain IMF set $\{c_{i}(t)\}_{i=1}^{M}.$
	\item[Step 3.] Repeat Step 1. and 2. for the signal realizations of WGN and estimate average IMF set
	\begin{equation*}
	\{\overline{c_{i}(t)}\}_{i=1}^{M}=\frac{1}{S}\Big( \{c_{i}(t)\}_{i=1}^{M}+\cdots +  \{c_{i}(t)\}_{i=1}^{M} \Big).
	\end{equation*}
\end{description}

\begin{figure}[!t]
	\centering
	\includegraphics[width=0.75\textwidth]{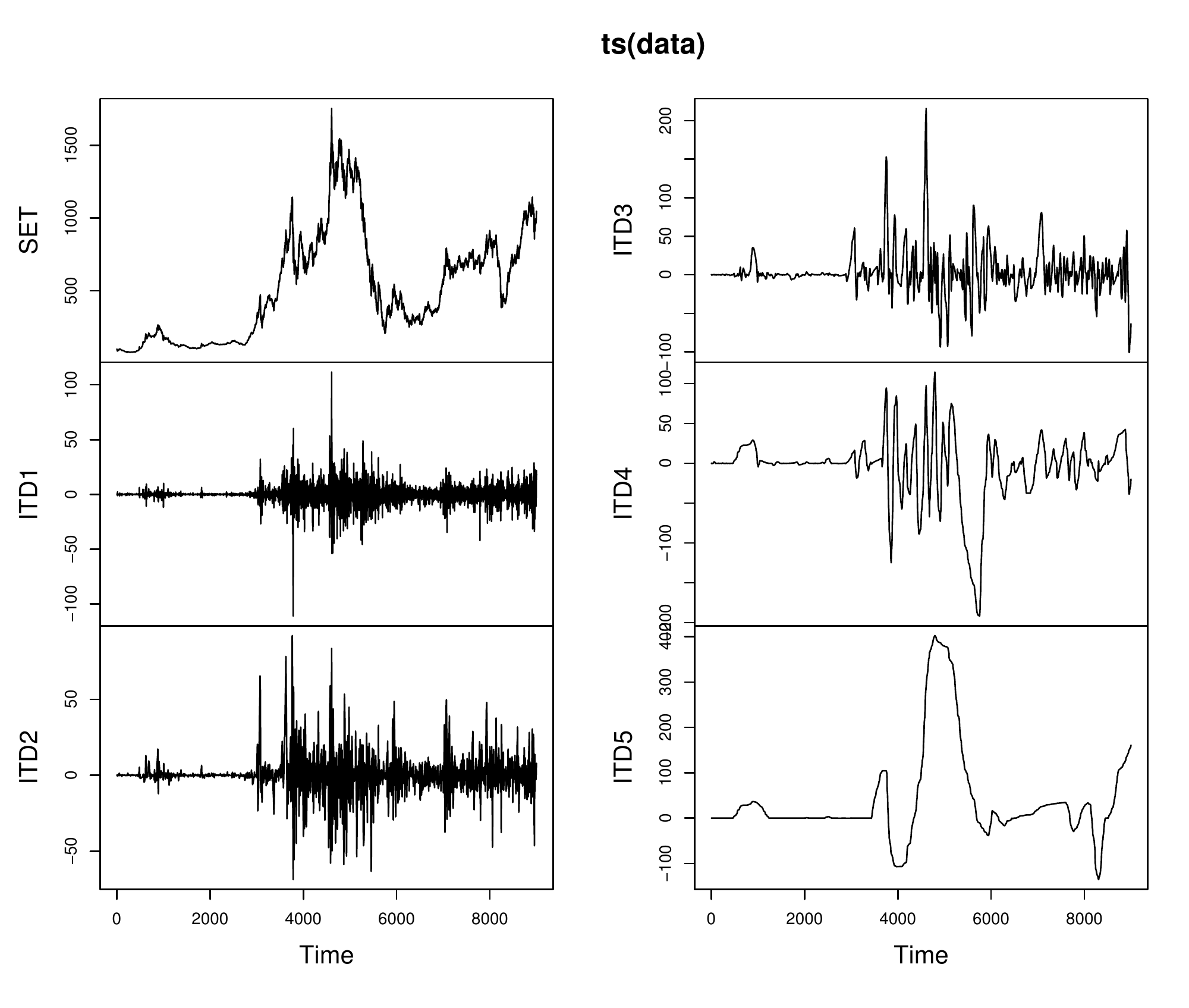}
	\caption{Intrinsic time scale decomposition $\mathrm{ITD}_{1}-\mathrm{ITD}_{5}$ of time series data of SET index daliy closed price. A data showned here is a daily closed price of SET index between 2/5/1975 to 12/09/2011 with 9000 time series data and $\mathrm{ITD}_{1}-\mathrm{ITD}_{5}$ at the same periods between raw data $9000$ data of SET index. \label{itd_set}}
\end{figure}

\subsection*{Intrinsic time scale decomposition}

Intrinsic time scale decomposition (ITD) \cite{Frei} decomposed the original signal into $\mathrm{ITD}_{i}(t)$ and monotonic trend $r_{1}(t)$ by using a baseline function
\begin{equation} 
x_{t}=\sum_{i=1}^{n}\mathrm{ITD}_{i}(t)+r_{1}(t),
\end{equation} 
with $\mathrm{ITD}_{i}(t)$ defined by using residua $H_{t}$ after the recursive substraction with baseline function $L_{t}$, similarly to EMD process,
\begin{equation*}  
x_{t}=L_{t}+H_{t}=Lx_{t}+(1-L)x_{t}.
\end{equation*}
The baseline function $L_{t}$ of $\mathrm{ITD}_{i}(t)$ is defined assuming
\begin{equation} 
L_{t}=L_{k}+\frac{L_{k+1} -L_{k}}{x_{k+1}-x_{k}}(x_{t}-x_{k}),
\end{equation}
where $L_{t}$ is an extremum location for $t\in (\tau_{k},\tau_{k+1}]$. The recursive process to decompose a time series is comming from the calculation of forward looking baseline function with three input parameters
\begin{itemize}
	\item the values of extrema $(x_{k},x_{k+1},x_{k+2})$,
	\item the locations of extrema $(\tau_{k},\tau_{k+1},\tau_{k+2})$,
	\item and the adjusting parameter $\alpha$
\end{itemize}
with
\begin{equation} 
L_{k+1}=\alpha\Big[  x_{k}+\frac{\tau_{k+1} -\tau_{k}}{\tau_{k+2}-\tau_{k}}(x_{k+2}-x_{k})\Big]  +(1-\alpha)x_{k+1} . 
\end{equation}

\begin{figure}[!t]
	\centering
	\includegraphics[width=0.9\textwidth]{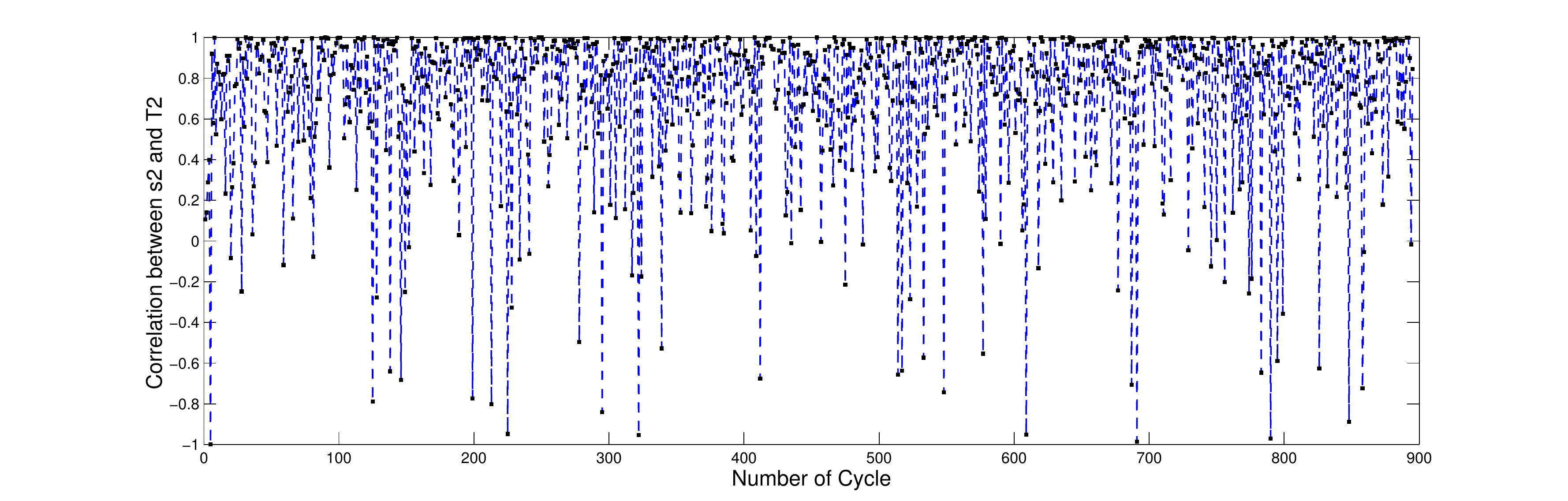}
	\caption{The correlation between $s_{2}$ and $T_{2}$ of $\ICHAIN$ of SET in the first $900$ cycles. \label{correlation}}
\end{figure}

\begin{figure}[!t]
	\centering
	\includegraphics[width=1\textwidth]{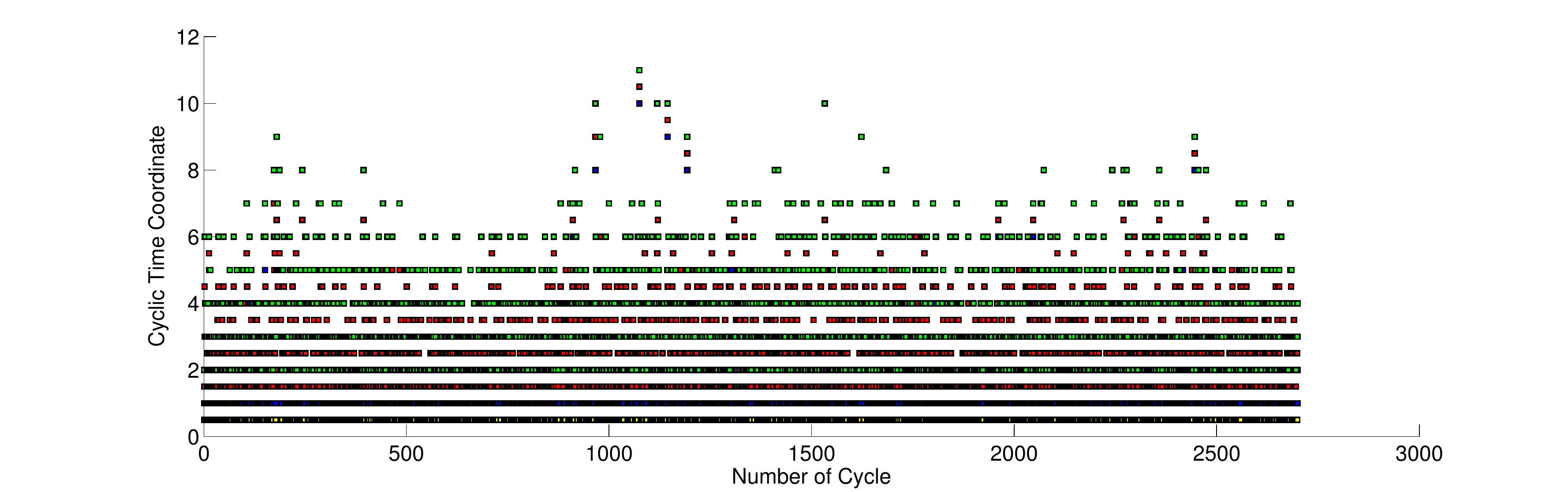} 
	\caption{The cyclic time coordinate of $\ICHAIN$ of SET daily closed price data from 2/5/1975 to 12/09/2011 with 9000 data point. The higher dot on the verticle line is $T_{4}$ cyclic time scale coordinate. It is a time from minimum to a next minimum point (labeled in green color). The lower dot on vertical line is $T_{1}$, a cyclic time scale from local minimum point to monotone function up (labeled in yellow color). The $T_{2}$ is labeled with blue color, it is a cyclic time coordinate from minimum to maximum point. The $T_{3}$ is labeled with red color, it is a cyclic time coordinate from minimum to monotone function down. For each cycle there are four points in the verticle line. The horizontal line represents time circle. The higher ITD will contain lower circle of time scale. We plotted only 2792 cycles in \ICHAIN with zeros crossing of cyclic coordinate for time series data on the leftside of panel. The highest peak label represent about $12$ days of cycle (labeled in blue color). \label{cyclic2}}
\end{figure}

\subsection*{$\ICHAIN$}
The EEMD transformation has a problem of the location of local extrema point of original signal if the location of input time series data and the height of IMF is not the same as the height of the original time series data. We solved this problem by performing the ITD, then sending the result of $\mathrm{ITD}_{1}$ to  EEMD transformation and to get IMF without the mentioned problem above. The result is called running $\ICHAIN$ with
\begin{equation} 
x_{t}=\sum_{i=1}^{n}\mathrm{ITD}_{i}(t)+r_{1}(t).
\end{equation} 
We select only $\mathrm{ITD}_{1}(t)$ to perform a further EEMD process with
\begin{equation} 
\mathrm{ITD}_{1}(t)=\sum_{i=1}^{n}c_{i}(t)+r_{2}(t),
\end{equation} 
where we call
\begin{equation} 
\ICHAIN(t)=c_{1}(t). \label{eq:ichain}
\end{equation} 

\begin{figure}[!t]
	\centering
	\includegraphics[width=0.3\textwidth]{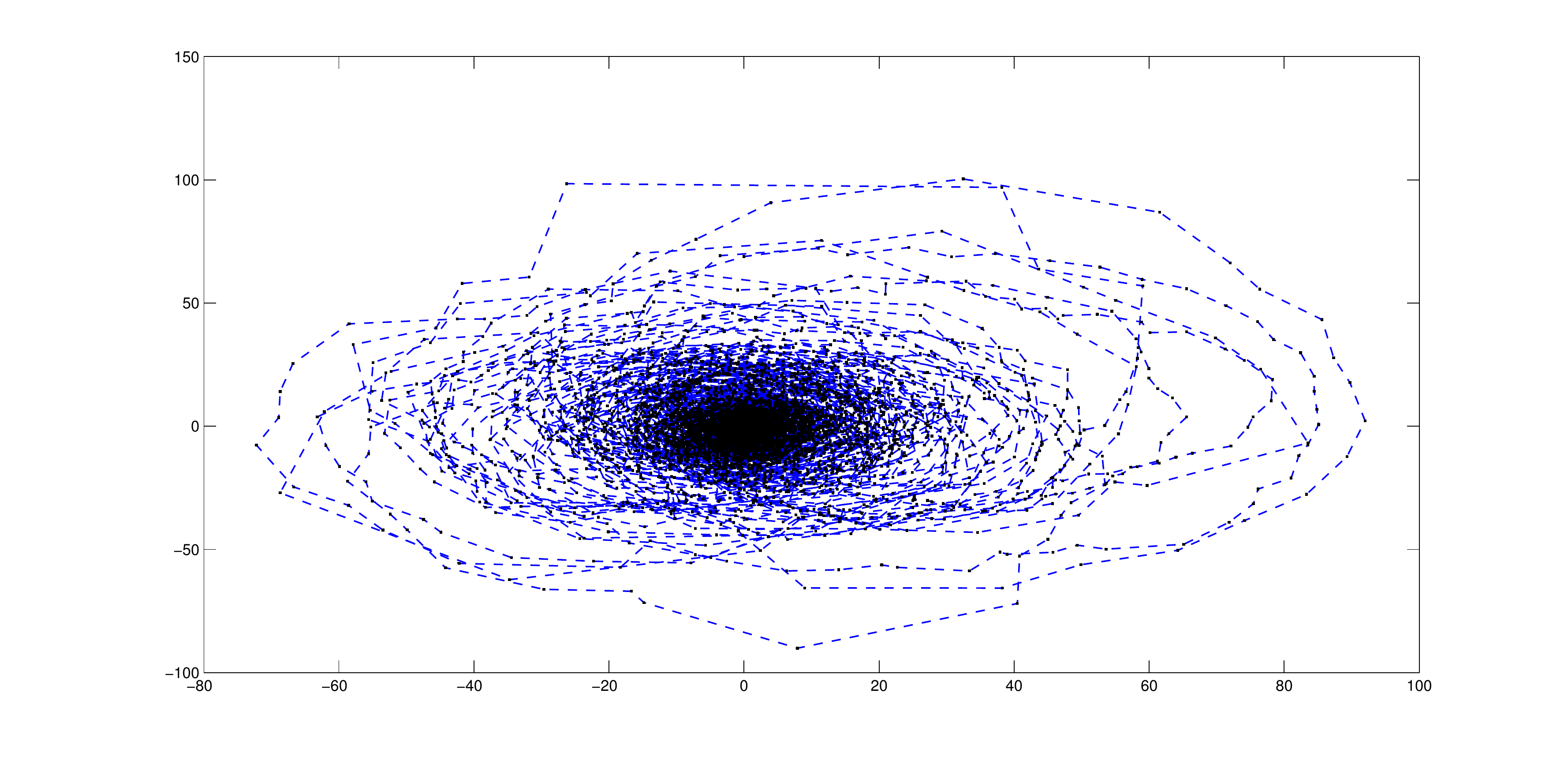}
	\includegraphics[width=0.3\textwidth]{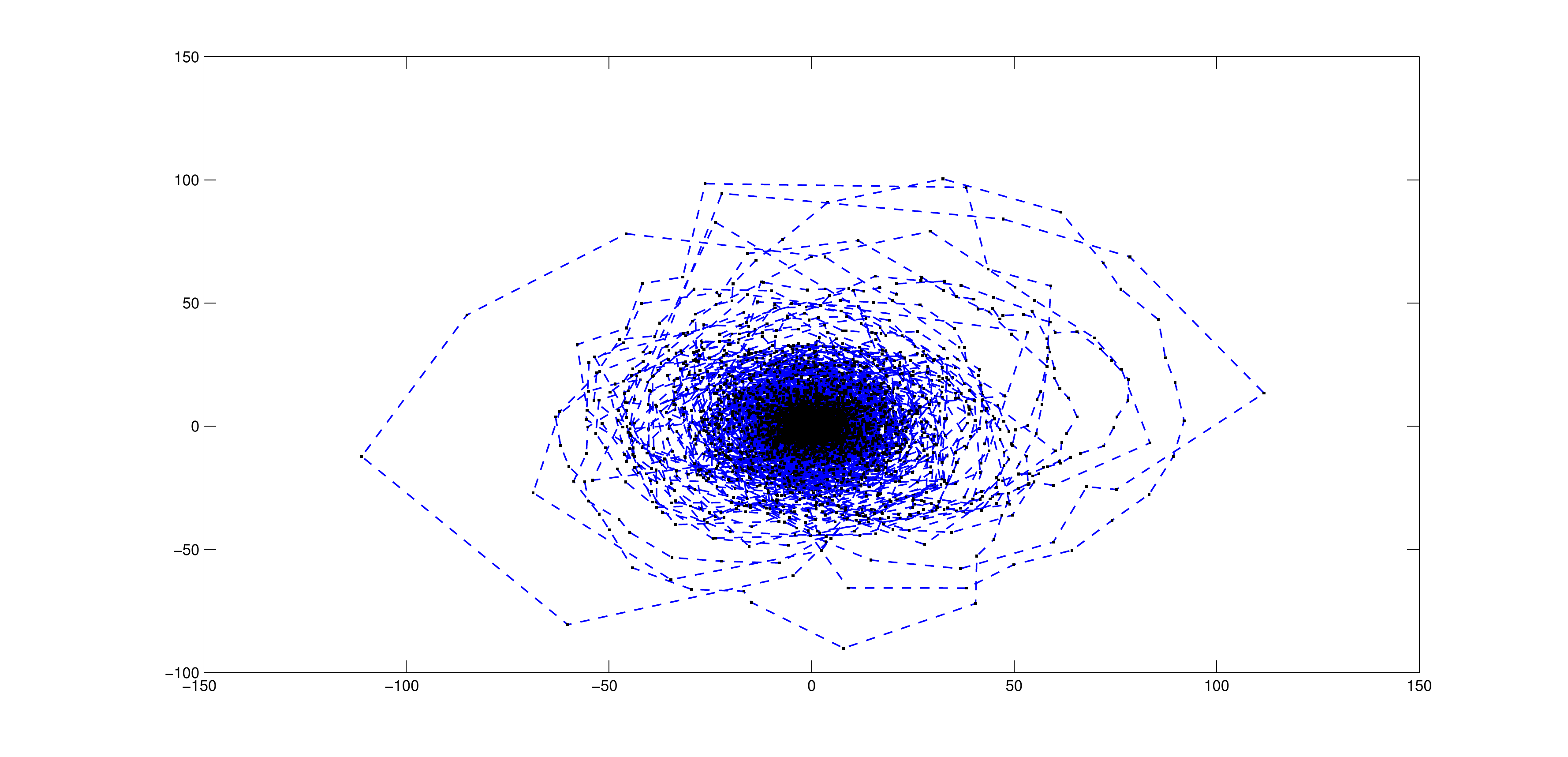}
	\includegraphics[width=0.3\textwidth]{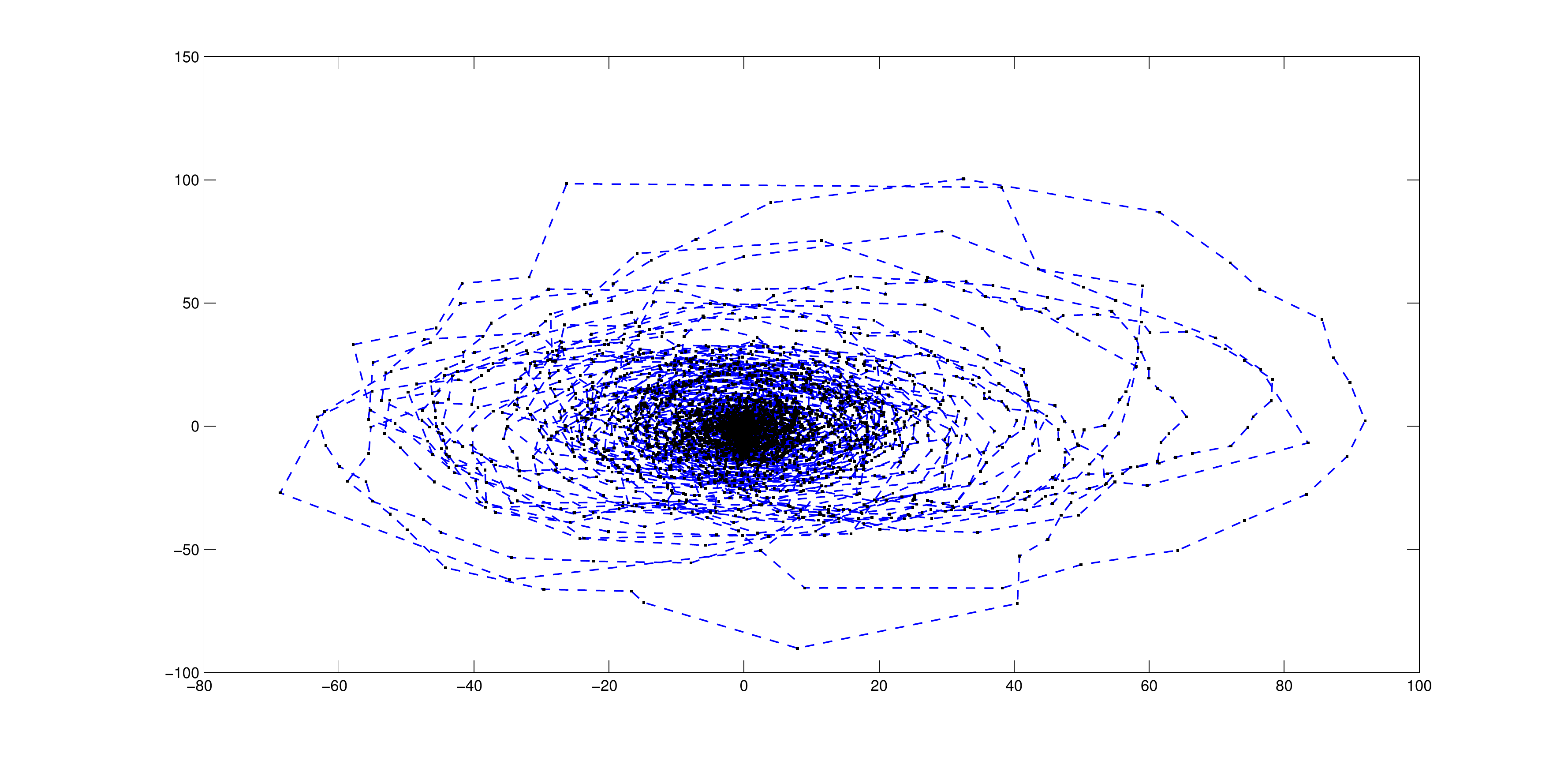}
	\caption{The graph of Hilbert transformation of $(ITD-IMF)chain(1)$ on the left, $\mathrm{ITD}_{1}$ in the middle and $\mathrm{ITD}_{2}$ on the right. \label{hilbert_itdimf1}}
\end{figure}

\begin{figure}[!t]
	\centering
	\includegraphics[width=0.3\textwidth]{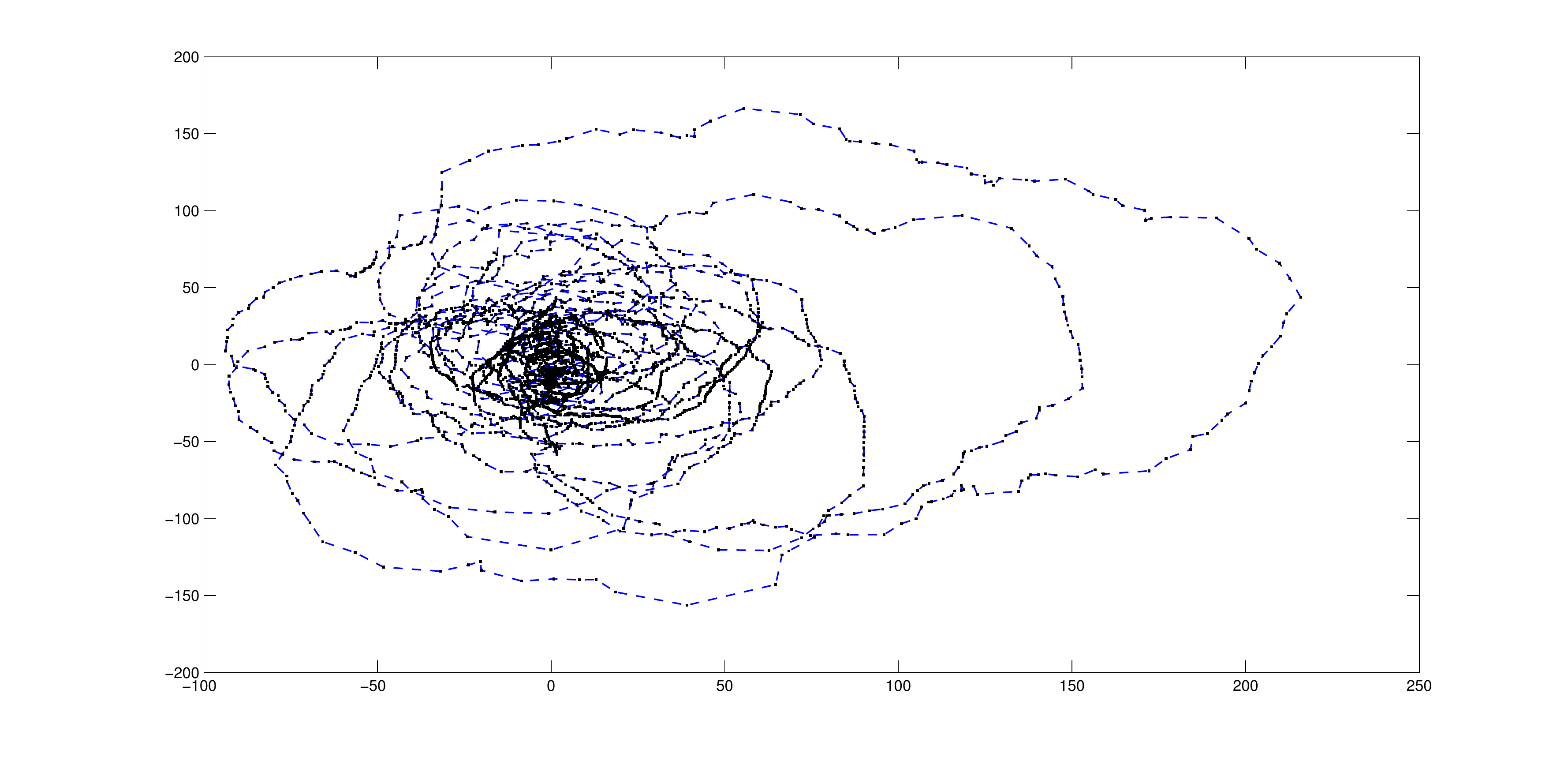}
	\includegraphics[width=0.3\textwidth]{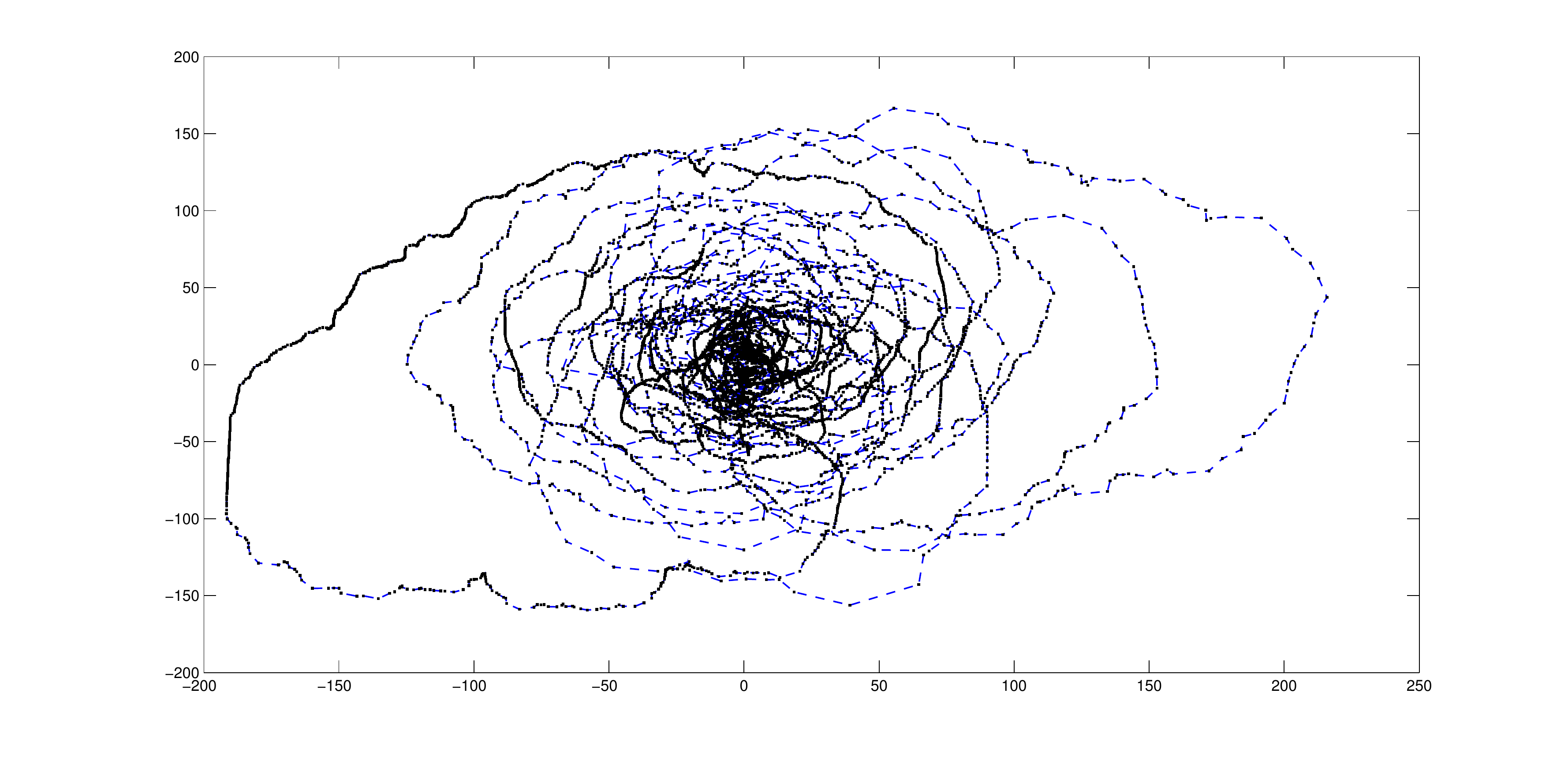}
	\includegraphics[width=0.3\textwidth]{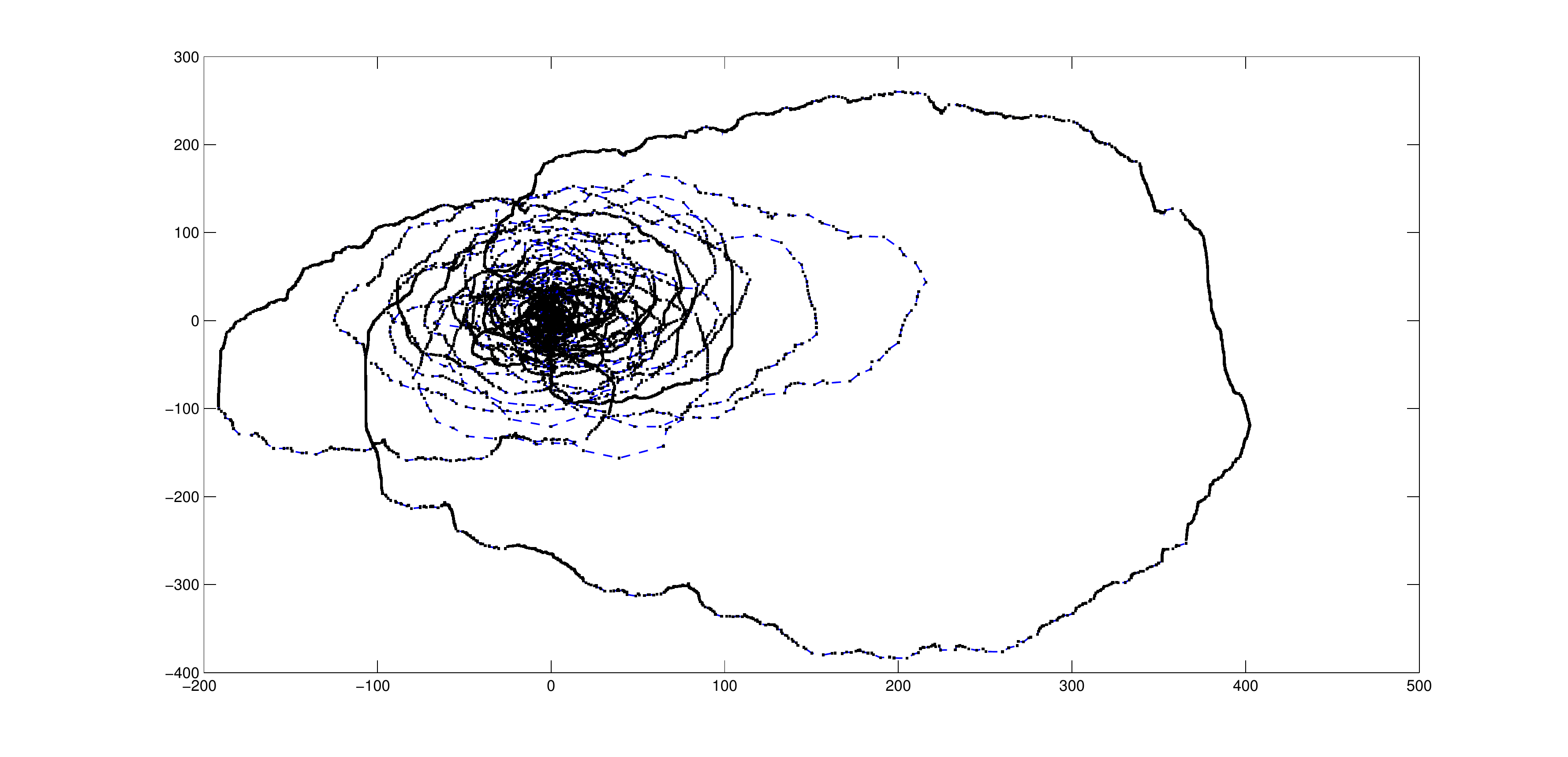}
	\caption{The graph of Hilbert transformation of $\mathrm{ITD}_{3}$ on the left, $\mathrm{ITD}_{4}$ in the middle and $\mathrm{ITD}_{5}$ on the right. \label{hilbert_itdimf2}}
\end{figure}

\begin{figure}[!t]
	\centering
	\includegraphics[width=0.3\textwidth]{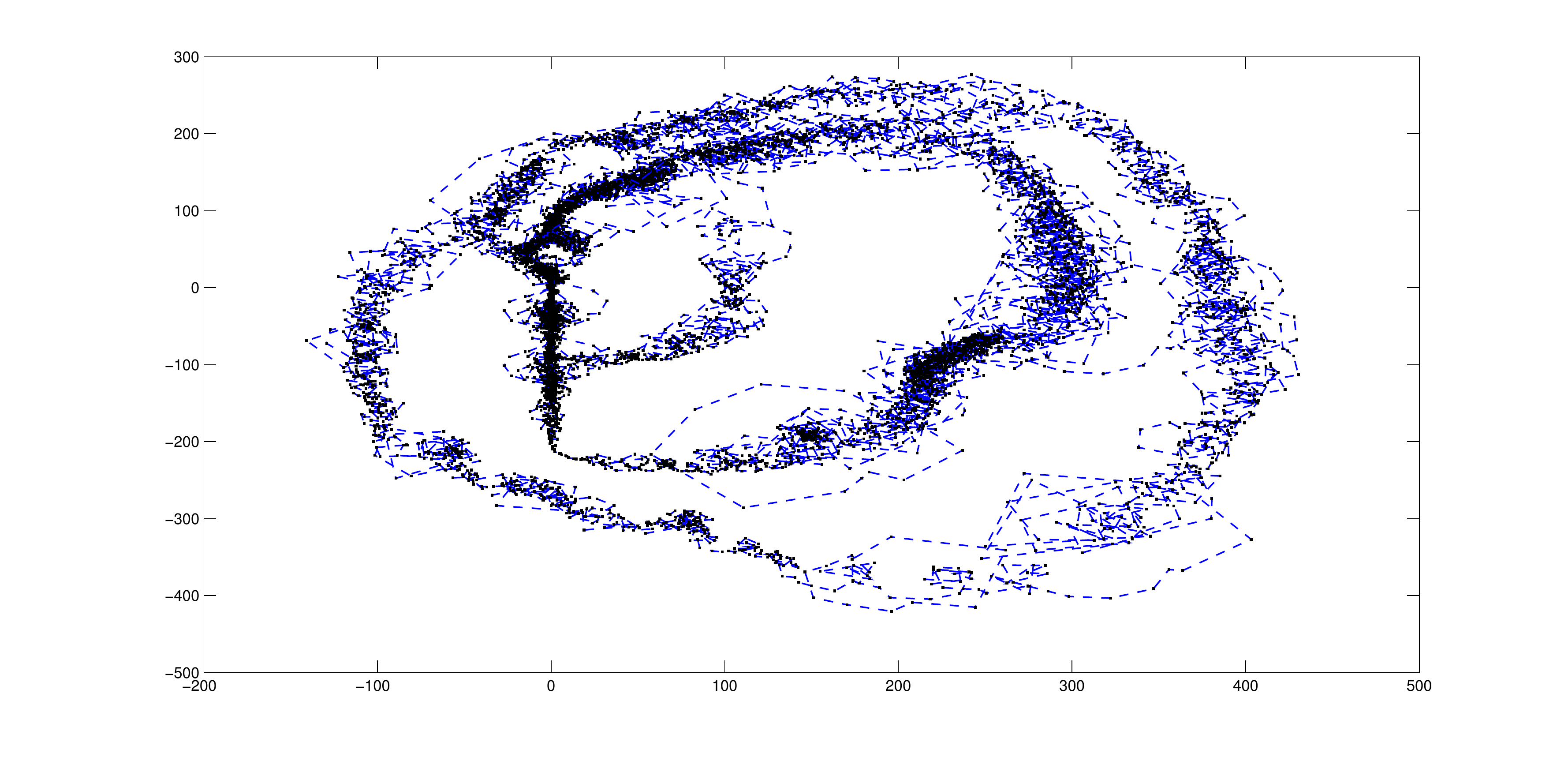}
	\includegraphics[width=0.3\textwidth]{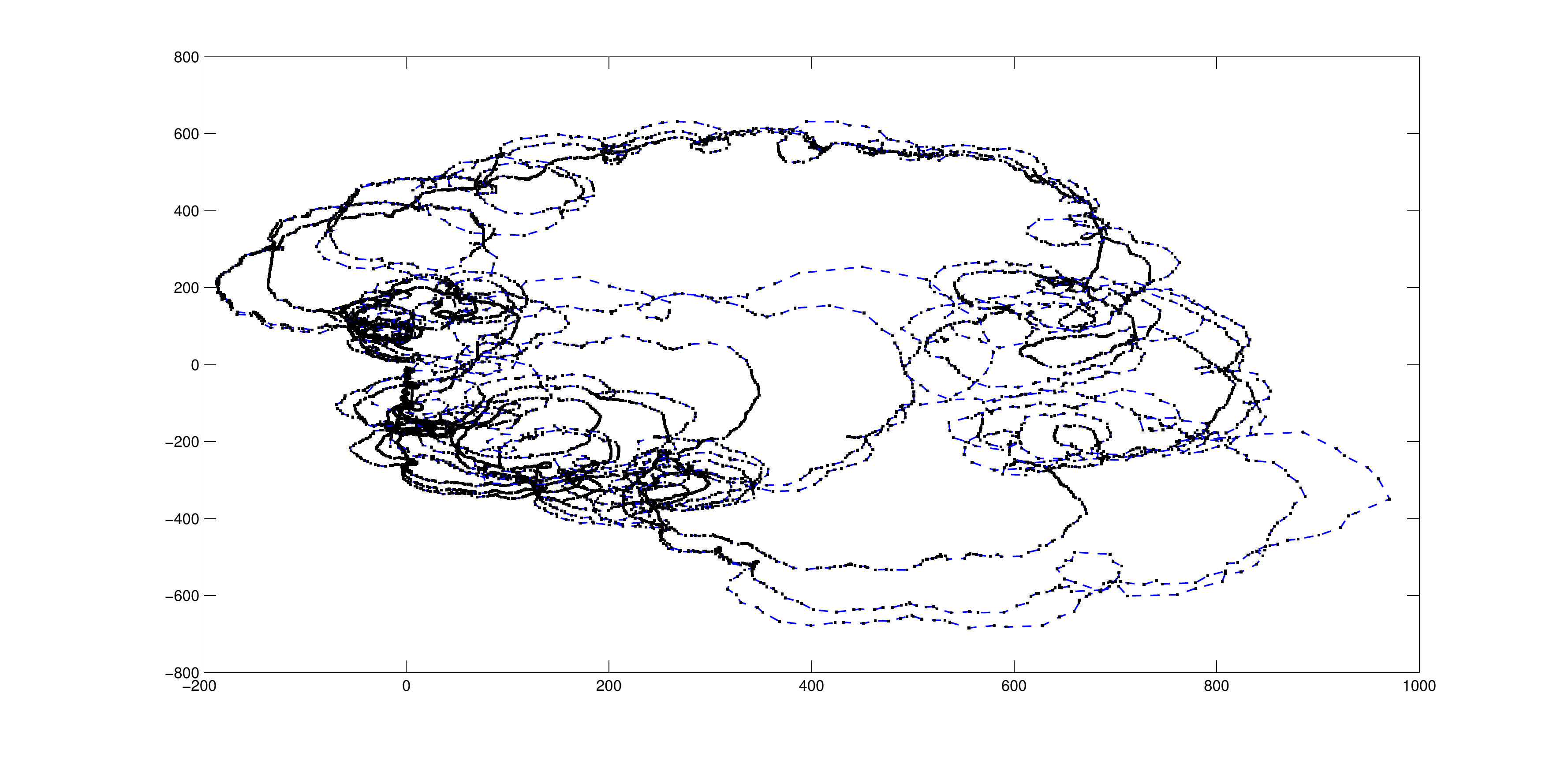}
	\includegraphics[width=0.3\textwidth]{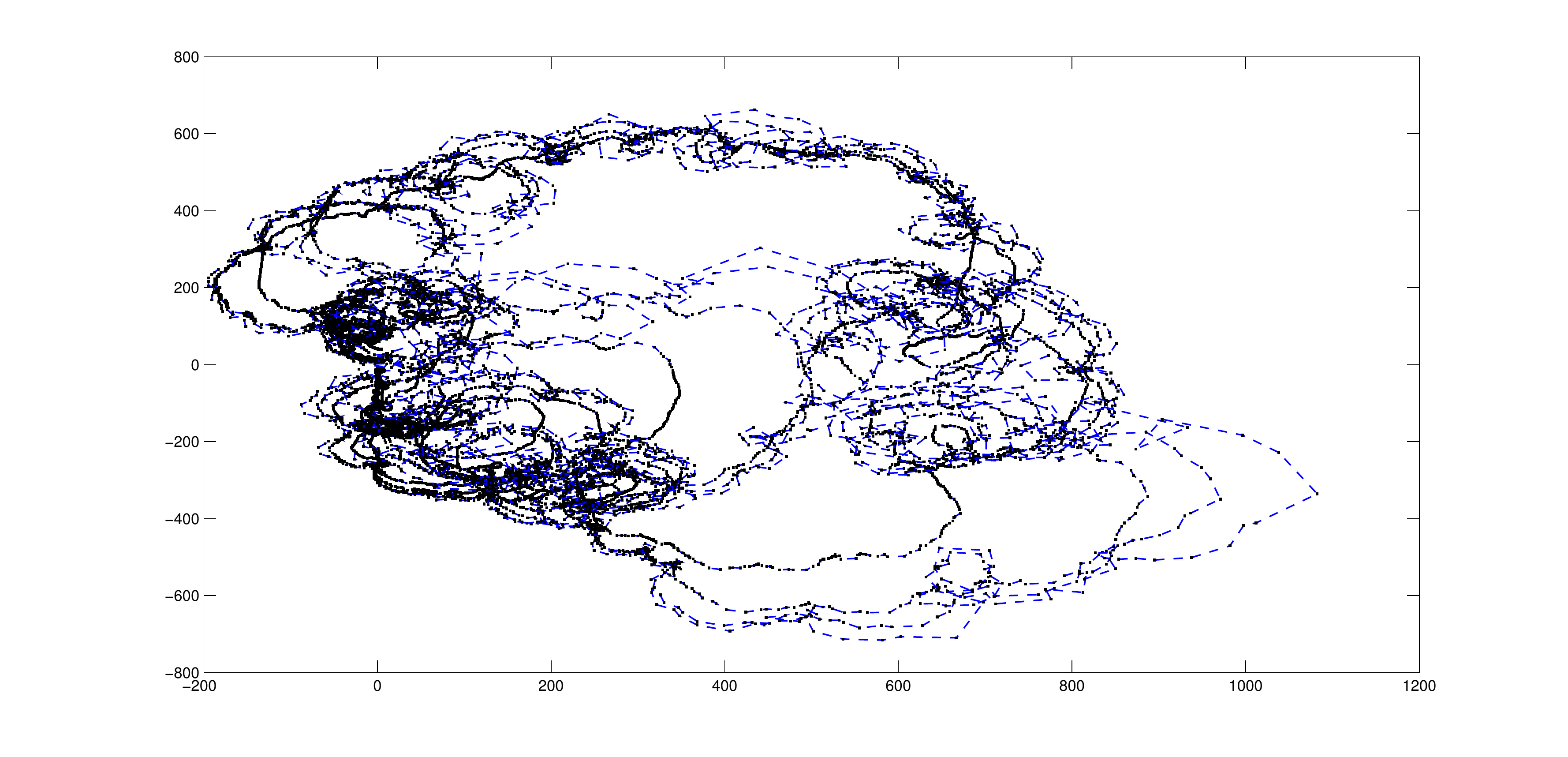}
	\caption{The graph of Hilbert transformation of $\mathrm{ITD}_{1}$ plus $\mathrm{ITD}_{5}$ on the left, $\mathrm{ITD}_{2}$ plus $\mathrm{ITD}_{3}$, $\mathrm{ITD}_{4}$, $\mathrm{ITD}_{5}$ in the middle and $\mathrm{ITD}_{1}$ plus $\mathrm{ITD}_{2}$, $\mathrm{ITD}_{3}$, $\mathrm{ITD}_{4}$, $\mathrm{ITD}_{5}$ on the right. \label{hilbert_itdimf3}}
\end{figure}

The important point of this result is a minimum structure of local maximum and local minimum state of financial time series data in which we can identify the minimum local structures of physiology of financial time series. The local maximum and minimum point of $\ICHAIN(t)$ lay at the same location of original signal $x_{t}$ without intermittency problem. If we can predict the local maximum and minimum state of $\ICHAIN(t)$ it means we can overcome the prediction of local maximum and local minimum state of original time series data. In this research we use $\ICHAIN(t)$ as skeleton of time series.
\begin{Definition}\label{itdchain}
	We call $\ICHAIN(t)$  of time series $x_{t}$ a skeleton of time series data $x_{t}$.
\end{Definition}

In this research, we have used financial time series data of daily closed price of SET, a Thai stock market index. We have mainly considered the daily closing prices of SET during the periods from the begining of market on 2/5/1975 to 12/9/2011, totally of $9000$ data points of time series for our data analysis. At first we have computed $\mathrm{ITD}_{1}-\mathrm{ITD}_{5}$ of SET index and result is shown in Fig.~(\ref{itd_set}). At second we have used the result of $\mathrm{ITD}_{1}$ for the computation of EEMD in order to get $\ICHAIN$. The EEMD computation was performed with the standard deviation of noise $0.05$ with $1000$ rounds of running an it has taken about $2$ hrs. per sample point.

\subsection*{Empirical analysis of cyclic time scale}

We have used data of $9000$ daily closed prices of SET. We have got only $2792$ cycles in $\ICHAIN$ with zeros crossing of cyclic coordinate for time series data. That means we have $2792$ maximum points which equal to minimum points of time series data. We have computed the correlation between state $s_{2}$ of $\ICHAIN$ and cyclic time scale $T_{2}$ (see Fig.~\ref{correlation}. The highest period in one smallest cycle $T_{4}$ is 11 days. We can notice from the highest point of graph Fig.~\ref{cyclic2} in cyclic time coordinate. 

The graphs of  Hilbert transformation of $\ICHAIN$ and $\mathrm{ITD}_{1}$ -- $\mathrm{ITD}_{5}$ of SET are shown in Figs.~\ref{hilbert_itdimf1}, \ref{hilbert_itdimf2}, \ref{hilbert_itdimf3}.

\subsection{Grey model}

Grey system theory is an interdisciplinary scientific area that was introduced in early 1980s by Deng \cite{deng}. Grey models require only a limited amount of data to estimate the behavior of unknown systems with its ability to deal with the systems that have partially unknown parameters. Grey models predict the future values of time series based only on a set of the most recent data depending on window size of the predictor. $\mathrm{GM}(1,1)$ type of Grey model is the most widely used in the literature. The differential equations 
\begin{equation} 
\frac{d\mathbf{x}^{(1)}}{dt}+a\mathbf{x}^{(1)}=b
\end{equation} 
have time varying coefficient $(a(t),b(t))$. Lets consider a time series sequence $\mathbf{x}^{(0)}=\{x^{(0)}(1),x^{(0)}(2),\cdots,x^{(0)}(n)  \},n\in \mathbb{N}$, that denotes the close  price of set whose accumulating operator (AGO) series is $\mathbf{x}^{(1)}=\{x^{(1)}(1),x^{(1)}(2),\cdots,x^{(1)}(n)  \},n\in \mathbb{N}$, where  
\begin{equation} 
x^{(1)}(t)=\sum_{i=1}^{t}x^{(0)}(i).
\end{equation}
In the above differential equation, $[a \quad b]^T$ is a sequence of parameters that can be found as
\begin{equation} 
[a\quad b]^{T}=(B^{T}B)^{-1}B^{T}Y
\end{equation} 
where
\begin{equation} 
Y=\{ x^{(0)}(2),x^{(0)}(3),\cdots, x^{(0)}(n)  \}
\end{equation} 
and
\begin{equation} 
B=\left [
\begin{array}{cc}
-\frac{x^{(1)}(1)+x^{(1)}(2) }{2} &1 \\
-\frac{x^{(1)}(2)+x^{(1)}(3) }{2} &1 \\
-\frac{x^{(1)}(3)+x^{(1)}(4) }{2} &1 \\
\vdots &\vdots\\
-\frac{x^{(1)}(n-1)+x^{(1)}(n) }{2} &1 \\
\end{array}
\right ]
\end{equation}

\begin{figure}[!t]
	\centering
	\includegraphics[width=0.8\textwidth]{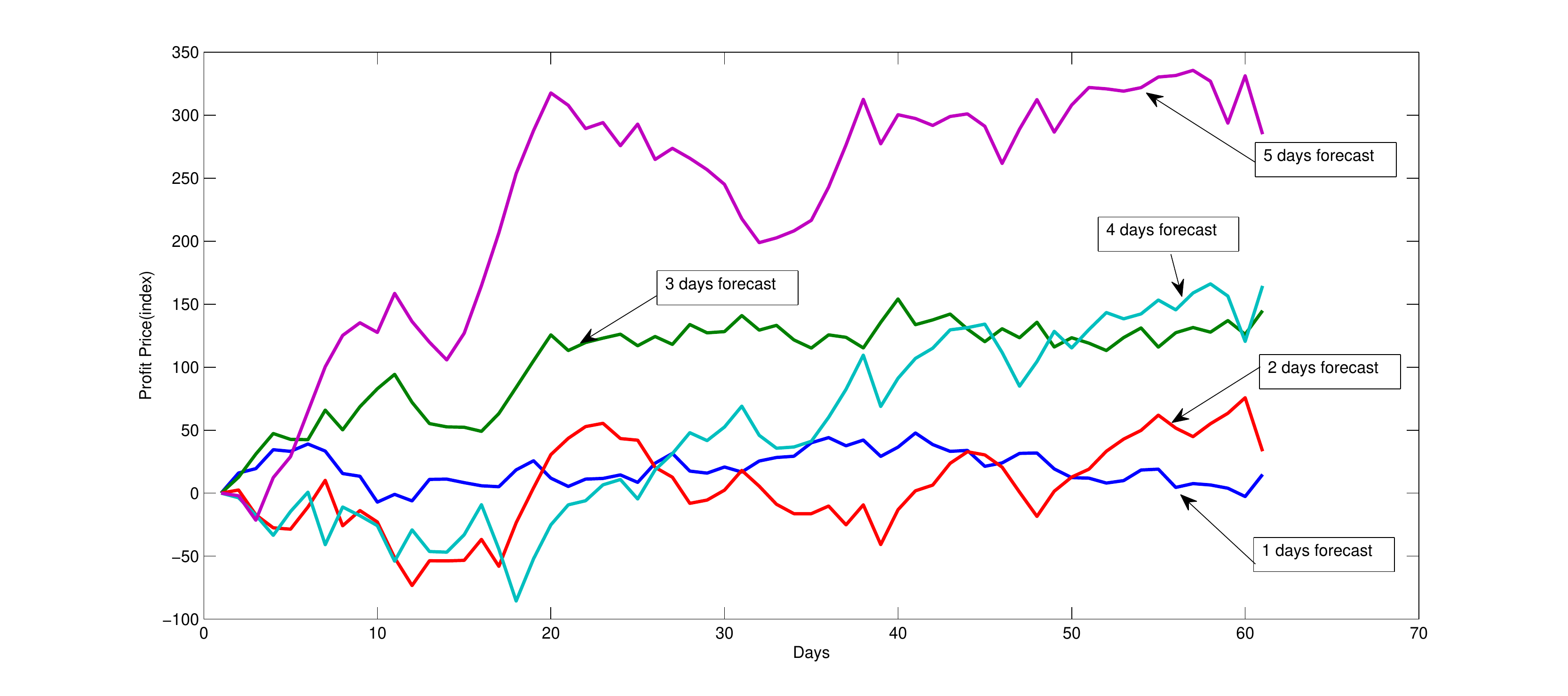}
	\caption{The performance plot of $5$ days forecast with holding a position in five days as the best result. The second rank is $4$ days forecast with holding a position within $4$ days. The worst case is one day forcast.\label{performance}}
\end{figure}

The solution of
\begin{equation} 
\hat{x}^{(0)}(t+1)=\hat{x}^{(1)}(t+1)-\hat{x}^{(1)}(t)
\end{equation} 
and
\begin{equation} 
\hat{x}^{(1)}(t+1)=[x^{(1)}(0)-\frac{b}{a}]e^{-at}+\frac{b}{a}
\end{equation} 
since by definition of AGO $x^{(1)}(0)=x^{(0)}(1)$.
Therefore
\begin{equation}
\hat{x}^{(0)}(t+1)=[x^{(0)}(1)-\frac{b}{a}]e^{-at}(1-e^{a})
\end{equation}
and the predicted value of the primitive data at time $(t+H)$ is
\begin{equation}
\hat{x}^{(0)}(t+H)=[x^{(0)}(1)-\frac{b}{a}]e^{-a(t+H-1)}(1-e^{a}).
\end{equation}

According the paper \cite{endeffect}, the improved $\mathrm{GM}(1,1)$ model was proposed, which has the architecture of GM-HHT. The equation
\begin{equation}
\frac{d\mathbf{x}^{(1)}}{dt}+a\mathbf{x}^{(1)}=b
\end{equation}
can be obtained from theorem below.

\begin{figure}[!t]
	\centering
	\includegraphics[width=0.65\textwidth]{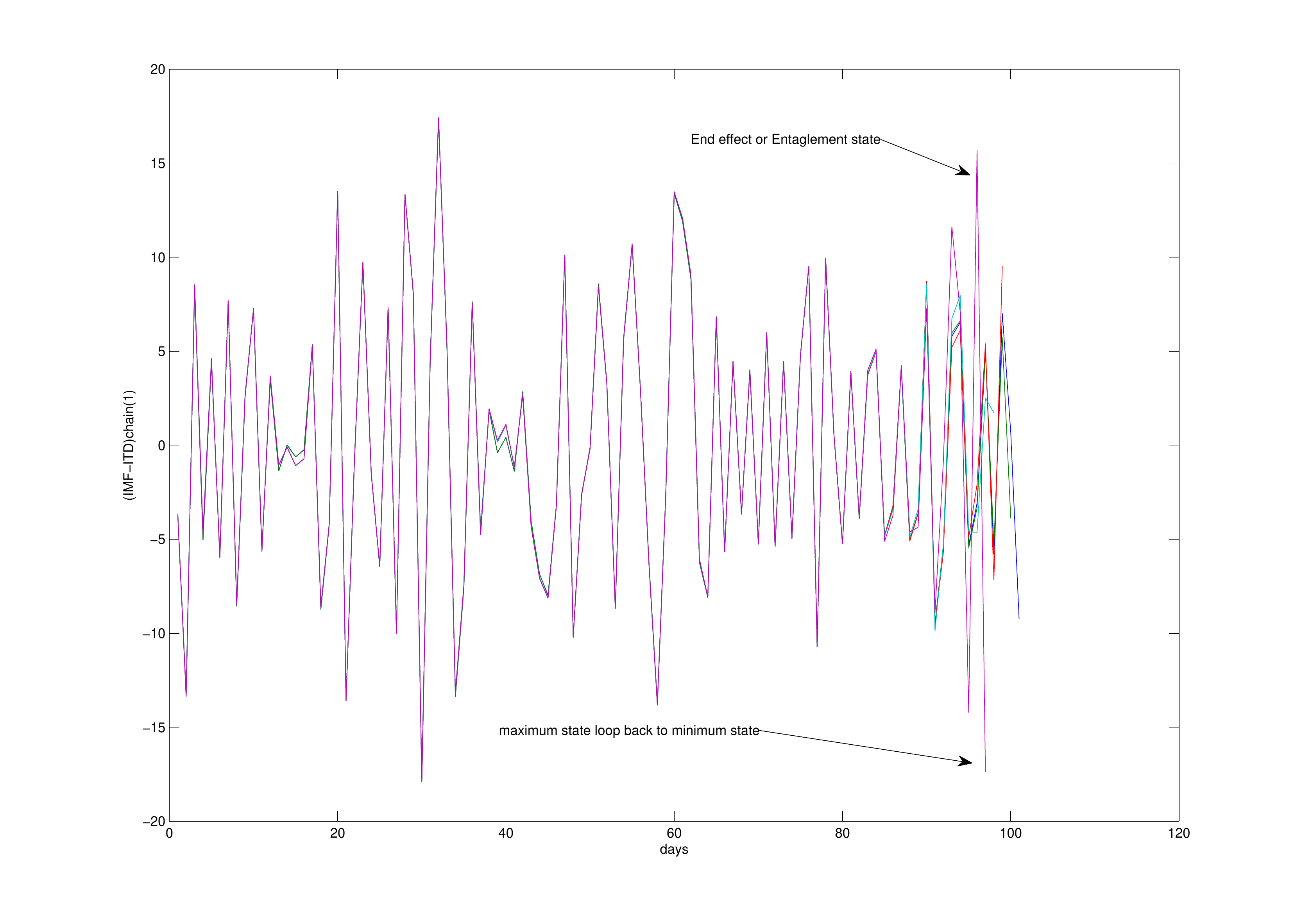}
	\caption{The end effect with entanglement state of loopback between maximum and minimum states. It can be used to forecast the maximum state one day ahead.\label{entaglement_emd}}
\end{figure}

\begin{Theorem}
Let the original signal $\mathbf{x}^{(0)}=\{ x^{(0)},\cdots ,x^{(n)} \}$. The discretize of the first order derivative of $\mathrm{GM}(1,1)$ can be obtained from 
\begin{equation}
	\mathbf{M'}=\left[   
	\begin{array}{c}
	m_{1}\\
	m_{1}\\
	\vdots\\
	m_{m}\\
	\end{array}
	\right ]
	=
	\left[   
	\begin{array}{c}
	b-ax^{(1)}(1)\\
	A^{-1}G\\
	b-ax^{(1)}(n)\\
	\end{array}
	\right ] 
\end{equation}
where $m_{i}=\frac{d\mathbf{x}^{i}}{dt}|_{t=i}$ and
\begin{equation}
	A=
	\left[   
	\begin{array}{cccccc}
	2 &0.5 &0 &\cdots &\cdots & 0\\
	0.5 &2& 0.5 & \cdots & 0& \vdots\\
	\vdots & \cdots & \cdots & \cdots &\cdots &\vdots\\
	\vdots & \cdots & \cdots & 0.5 &2 &0.5\\
	\vdots & \cdots & \cdots & 0 &0.5 &2\\
	\end{array}
	\right ] 
\end{equation}
\begin{equation}G=
	\left[   
	\begin{array}{c}
	g_{2}-0.5(b-ax^{(1)}(1))\\
	g_{3}\\
	g_{4}\\
	\vdots\\
	g_{n-2}\\
	g_{n-1}-0.5(b-ax^{(1)}(n))\\
	\end{array}
	\right ] 
\end{equation}
where
\begin{equation}
	g_{j}=1.5(x^{(0)}(j)+ x^{(0)}(j+1)),j=2,3,\cdots, n.
\end{equation}
We can write the equation as
\begin{equation}
	\mathbf{M}'+a\mathbf{x}^{(1)}=b.
\end{equation}
\end{Theorem}
\emph{Proof:} see \cite{endeffect}.

\subsection{Performance test of directional prediction of SET index}

The five days ahead forecast by using of improved $\mathrm{GM}(1,1)$-ITD-HHT-ANN is performed with $92$ data set of out of sample test. The test data start from 2/3/2014 with data number $9528$ to 24/6/2014 with data number $9622$. $9622$ means SET index of daily closed price of date number $9622$ since market started.
The result of calculation was used for the performance test of profit of short and long positions of IndexFutures Market. The result of the performance test of our prediction model is test over $60$ sample data. The trading day start with data number $9528$ to data number $9588$. The graph of performance test is shown in Figs.~\ref{performance}. For one day ahead forecast we can notice only which state is maximum state $s_{2}$ in our time series data. Then we can open a short position from that state. We obtain average accuracy of this method by average all $5$ days prediction in $92$ days of out of sample test. We get
average	accuracy at	$51.96 \%$ with standard deviation (SD) at $0.501430242$.

In this work we also detected entanglement state of time series data after using $\ICHAIN$ transformation. We found that mostly the entanglement state occurred, when time series data is in the maximum state $s_{2}$.

The picture of down direction of stock index can be notice one dayahead in out of sameple test within modeling of entanglement state in time series data of our empirical analysis is analogy with en deffect of our $\ICHAIN$, shown in Fig.~\ref{entaglement_emd}.

\end{document}